\documentclass[final]{ws-p10x7}

\usepackage{fleqn}

 \usepackage[dvips]{graphicx}

\input mystyle.sum

\input cleo-rep.defs

\newcommand{\aerr}[3]   {\mbox{${{#1}^{+ #2}_{- #3}}$}}
\newcommand{\derr}[5]   {\mbox{${{#1}^{+ #2}_{- #3}{}^{+ #4}_{- #5}}$}}

\newlength{\Figwidth}
\def\arXiv{arXiv}
\def\Where{arXiv}

\ifx\Where\arXiv
\catcode`\@=11 
\def\cropmarks{ }
\def\ps@myheadings{ }
\def\ps@headings{ }
\def\@evenhead{ }
\def\@oddhead{ }
\def\@oddfoot{{\parbox{\textwidth}{\vspace*{4ex}
\centerline{To be published in the}
\centerline{\it Proceedings of the XX International Symposium on}
\centerline{\it Lepton and Photon Interactions at High Energies}
\centerline{Rome, July 23 -- 28, 2001}}}}
\def\@evenfoot{ }
\catcode`\@=12 
\fi

\begin{document}

\title{\ifx\Where\arXiv \hfill\textmd{\normalsize CLNS 01/1768}\\ \fi
CLEO Results: \boldmath{$B$} Decays\ifx\Where\arXiv\\[-2ex]\else\\[1ex]\fi}
\author{David G. Cassel}
\address{Newman Laboratory, Cornell University, Ithaca, NY 14853, USA\\
E-mail: dgc@lns.cornell.edu}

\twocolumn[\maketitle\abstract{Measurements of many Standard Model constants
are clouded by 
uncertainties in nonperturbative QCD parameters that relate measurable
quantities to the underlying parton-level processes.  Generally these
QCD parameters have been obtained from model calculations with large
uncertainties that are difficult to quantify.  The CLEO Collaboration
has taken a major step towards reducing these uncertainties in
determining the CKM matrix elements $\Vcb$ and $\Vub$ using new
measurements of the branching fraction and photon energy spectrum of
$\btosgamma$ decays.   This report includes: the new CLEO measurements of
$\btosgamma$ decays,
$\Vcb$, and $\Vub$; the first results from \Cleoiii\ data -- studies of $B \to K\pi$,
$\pi\pi$, and $K\Kbar$ decays; mention of some other recent CLEO $B$ decay results;
and plans for operating CESR and CLEO in the charm threshold region.}]

\maketitle

\section{Introduction}

New results from CLEO include measurements of:
\Begitem
\item the branching fraction and photon energy spectrum of  $\btosgamma$ decays,
\item $\Vcb$ from moments of hadronic mass in
$\BtoXclnu$ decays and photon energy in
$\btosgamma$ decays, 
\item $\Vub$ from the spectra of lepton momentum in $\BtoXulnu$ decays and
photon energy in $\btosgamma$ decays,
\item $\Vcb$ from $\BtoDstarlnu$ decay, and
\item branching fractions and upper limits for the charmless hadronic
decays $\BtoKpi, \pi\pi$, and $K\Kbar$ decays from \Cleoiii\  data.
\Enditem
This report includes these measurements, mention of some other recent CLEO results
and discussion of future plans for operating CESR and CLEO in the charm threshold
region. At this conference 
Belle~\cite{belle:rome} and {\sc BaBar}~\cite{babar:rome} also  presented 
experimental results on some of these topics, and many of the theoretical
issues were discussed by Neubert~\cite{neubert:rome},
Barbieri~\cite{barbieri:rome}, Wise~\cite{wise:rome}, and
Isidori~\cite{isidori:rome}.  

\Begfigure{htb}
\includegraphics[width=\Figwidth]{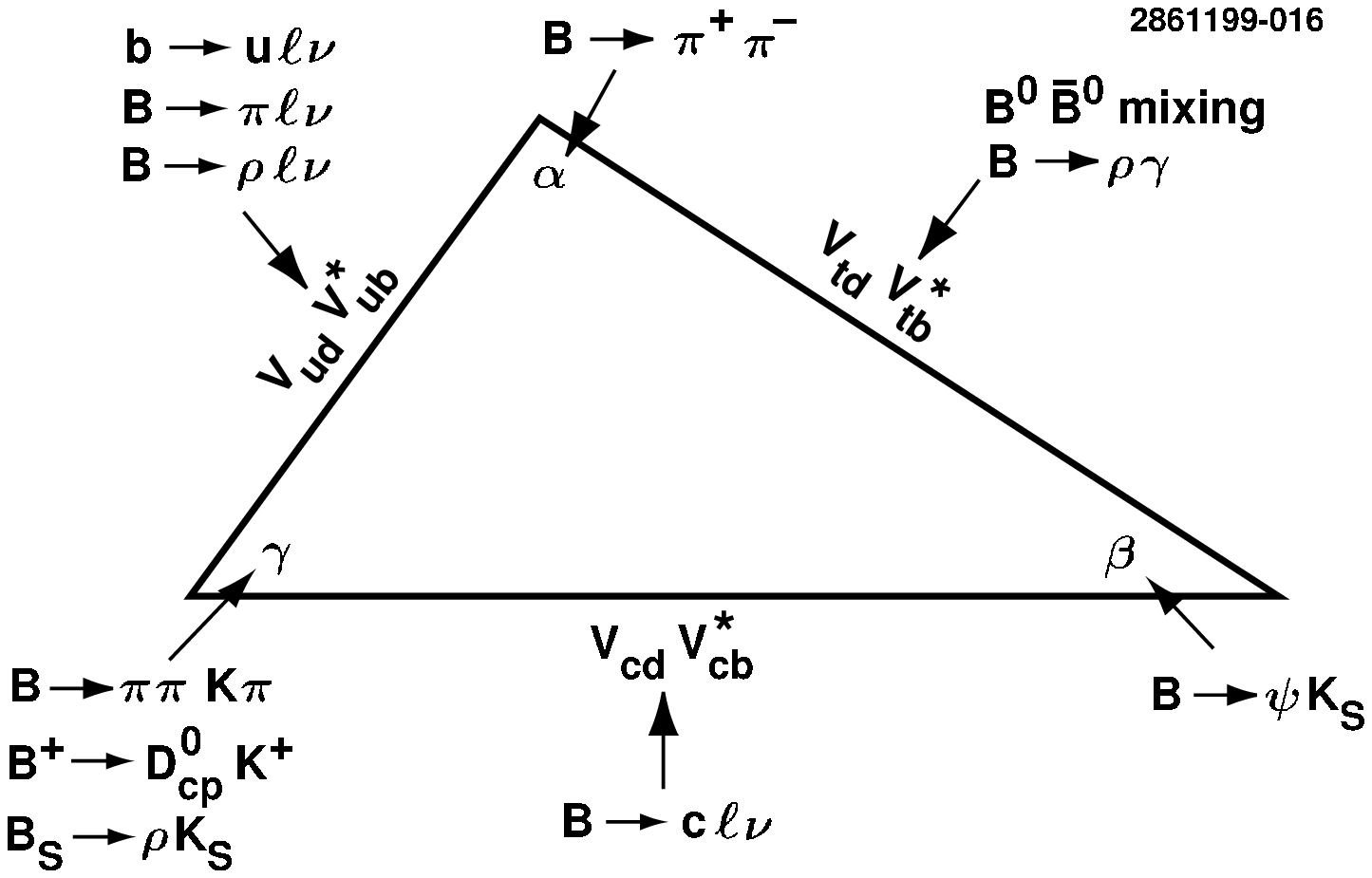}
\Endfigure{The unitarity triangle and some of the $B$ meson
measurements that can contribute to determining the angles and CKM matrix
elements.\vspace*{-4ex}}{fig:unitriangle}

Common goals of all $B$ physics programs include: identifying $B$ decay modes
and  accurately measuring $B$ branching fractions and the CKM matrix elements
$\Vcb$,
$\Vub$, $\Vtd$, and $\Vts$. 
\Fig{fig:unitriangle} illustrates the variety of the $B$ meson decays
that can contribute to measurements of CKM matrix elements and the unitarity
triangle.  However, the importance of $B$ decays arises from the possibility
that the key to understanding $CP$ violation can be found in the $b$ quark
sector.  Earlier today
BaBar~\cite{babar:beta} and Belle~\cite{belle:beta} reported major advances
in this direction -- statistically significant measurements of the
$CP$ violating parameter $\sin(2\beta)$, where the angle $\beta$ is illustrated in
\Fig{fig:unitriangle}.

According to conventional wisdom the amount of $CP$ violation
in the Standard Model (\ie, in the CKM matrix) is not sufficient to account for
the observed matter-antimatter asymmetry in the universe.  Hence, major goals of
$B$ physics programs also include searches for $CP$ violation and other New
Physics beyond the Standard Model (SM). Phenomena beyond the SM may appear in 
$B$ decays involving loops, such as rare charmless hadronic
$B$ decays and $\btosgamma$ decays.  Accurate measurement of
CKM matrix elements and searches for new physics beyond the SM are the principal
priorities of the CLEO $B$ physics program.

\Begfigure{htb}
\includegraphics[width=\Figwidth]{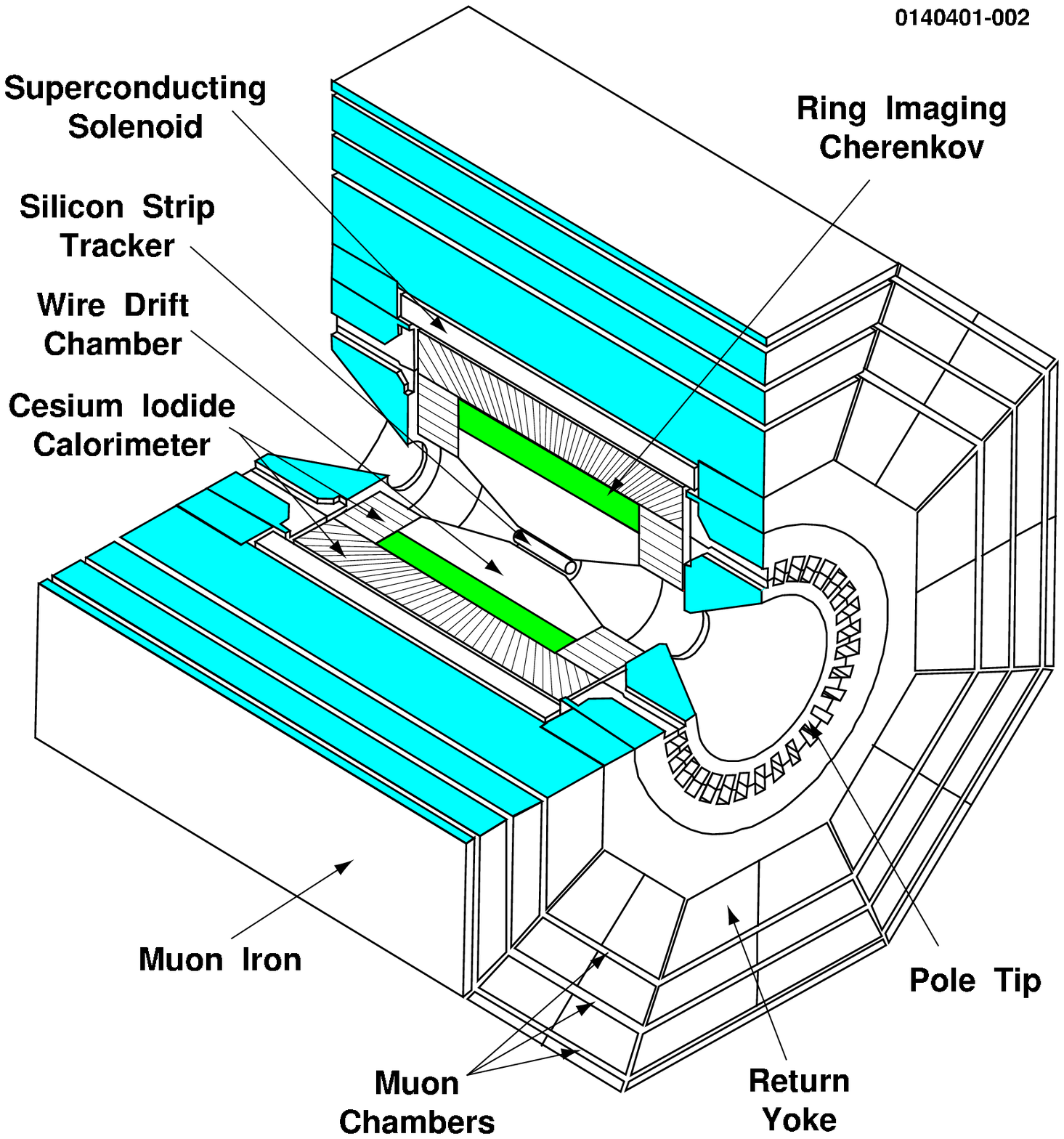}
\Endfigure{The \Cleoiii\ detector.}{fig:cleoiiidet}

\vspace*{-1ex}

\section{The CLEO Detectors and Data}

We obtained the $B$ decay results reported here using three configurations of the
CLEO detector, called \Cleoii, \Cleoiiv, and \Cleoiii. 
The \Cleoiii\ detector is illustrated in \Fig{fig:cleoiiidet}.
The CsI calorimeter, superconducting coil, magnet iron, and muon chambers
are common to all three detector configurations.  In the \Cleoiii\ upgrade,
the \Cleoiiv\  silicon vertex detector, drift chamber and time of flight counters
were replaced by a new silicon vertex detector, a new drift chamber, and a new Ring
Imaging Cherenkov detector, respectively.  \Tab{tab:cleoiiiperform} describes the 
performance achieved with the \Cleoiii\ detector.

\ifx\Where\arXiv \pagestyle{plain} \fi

\Begtable{t}{\Cleoiii\ detector performance.}{tab:cleoiiiperform}
\Begtabular{|l|l|}
Component & Performance \\
\hline
Tracking  & 93\% of $4\pi$; \\[-0.5ex]
          & at $p = 1$ GeV/$c$ \\[-0.5ex]
          & $\sigma_p/p=0.35$\%; \\[-0.5ex]
          & $dE/dx$ resolution 5.7\% \\[-0.5ex]
          & for minimum-ionizing $\pi$ \\
\hline
RICH      &  80\% of $4\pi$;  \\[-0.5ex]
          &  at $p=0.9$ GeV/$c$ \\[-0.5ex]
          &  87\% kaon efficiency with \\[-0.5ex]
          &  0.2\% pion fake rate \\
\hline
Calorimeter & 93\% of $4\pi$; $\sigma_E/E =$ \\[-0.5ex]
            & 2.2\% at $E = 1$ GeV \\[-0.5ex]
            & 4.0\% at $E = 0.1$ GeV \\
\hline
Muons     & 85\% of $4\pi$ for $p > 1$~GeV/$c$ \\
\hline
Trigger   & Fully pipelined; \\[-0.5ex]
          & Latency $\sim 2.5~\mu$s; \\[-0.5ex]
          & Based on track and \\[-0.5ex]
          & shower counter topology\\
\hline
DAQ   & Event Size: $\sim 25$~kByte; \\[-0.5ex]
          &  Throughput $\sim\, 6$~MB/s \\
\Endtabular
\Endtable


\Begtable{!hbt}{The numbers of $B\Bbar$ events recorded and the $\Upsilon(4S)$
and continuum integrated luminosities for the three CLEO
detector configurations.}{tab:cleodata}
\Begtabular{|lccc|}
Detector & ~~$\Upsilon(4S)$~~ & Cont. & $B\Bbar$\\[-0.75ex] 
         & \fbinv\ & \fbinv\ & ($10^6$) \\ \hline
CLEO II    &  ~3.1 & 1.6 &  ~3.3 \\
CLEO II.V  &  ~6.0 & 2.8 &  ~6.4 \\ \hline
Subtotal   &  ~9.1 & 4.4 &  ~9.7 \\ \hline
CLEO III   &  ~6.9 & 2.3 &  ~7.4 \\ \hline
Total      & 16.0 & 6.7 & 17.1 \\
\Endtabular
\Endtable


We accumulated a total of
17.1~M $B\Bbar$ events at the $\Upsilon(4S)$ and we devoted about 30\%
of our luminosity to running in the continuum just below the $\Upsilon(4S)$.
These continuum data were essential for determining backgrounds for the
inclusive measurements.  The breakdown of the data samples among the
different detectors, the $\Upsilon(4S)$, and the continuum are
summarized in \Tab{tab:cleodata}.
Only CLEO II data are used in the exclusive $\Bzbar \to \Dstarp \ellm \nuellbar$
analysis,  CLEO II and II.V data are now used in most other analyses, and CLEO III
data are used for the new $B \to K \pi,~\pi\pi$, and $K\Kbar$ results.

\section{\boldmath$\btosgamma$ Decays}
\label{sec:bsgamma}

The radiative penguin diagram illustrated in \Fig{fig:btosgamma} 
is responsible for radiative decays of $B$ mesons.  
The branching fraction, $\calB(\btosgamma)$, for inclusive $\BtoXsgamma$ decays is
sensitive to charged Higgs or other new physics
beyond the SM in the loop, and to anomalous $WW\gamma$ couplings.
Reliable QCD  calculations of 
$\calB(\btosgamma)$ in next to leading order (NLO) are available for comparison with
experimental measurements. On the other hand, exclusive $B \to K^{*(*)} \gamma$
branching fractions are sensitive to hadronization  effects and therefore
cannot be used in reliable searches for new physics.
\Begfigure{!htb}
\empenguin{$\Bbar$}{${b}$}{$\bar{q}$}{$s$}{$\bar{q}$}{$\bar{X}_s$}
\Endfigure{Radiative penguin diagram for $\Bbar \to \bar{X}_s \gamma$
decays.  The photon can couple to the $W$ or any of the other quarks.  The
observed hadronic final state $\bar{X}_s$ arises from the hadronization of the
$s$ and $\bar{q}$ quarks.}{fig:btosgamma}


Only \Cleoii\ data were available for the original CLEO
measurement~\cite{cleo:bsgamma1} of $\calB(\btosgamma)$. We now report an
update~\cite{cleo:bsgamma2} using the full
\Cleoii\ and \Cleoiiv\ data sample; a total of almost a factor of 3 more data than
were used in the earlier analysis.

The basic $\btosgamma$ signal is an isolated $\gamma$ with energy, 
$2.0 < E_\gamma < 2.7$ GeV.  This includes essentially all of the $E_\gamma$
spectrum. Previously CLEO used only the range $2.2 < E_\gamma < 2.7$ GeV.
There is much less model dependence in the new result since essentially the
entire $\btosgamma$ spectrum is now measured.\newpage

\Begfigure{htb}
\includegraphics[width=\Figwidth]{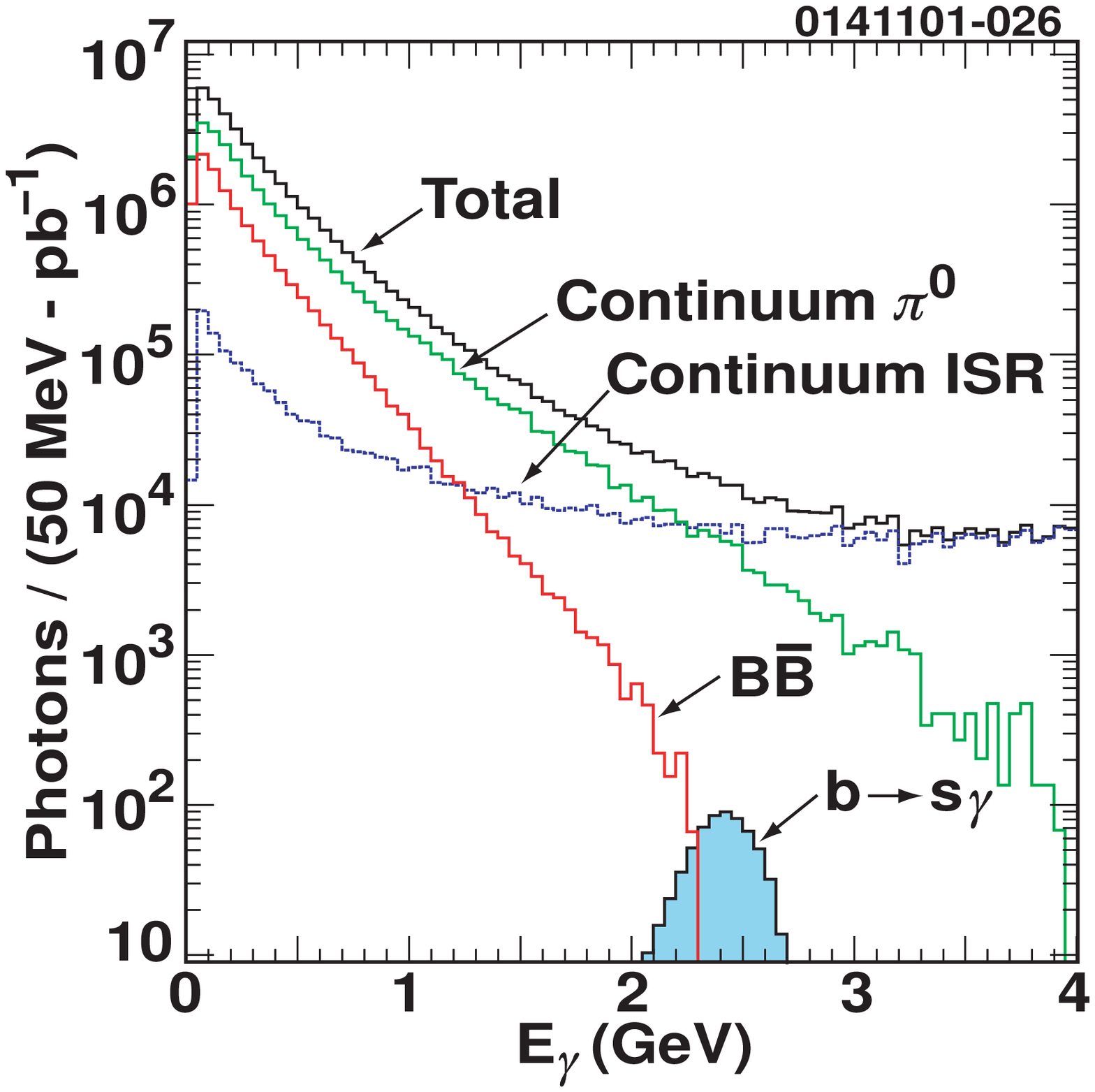}
\Endfigure{Monte Carlo estimate of the inclusive $\gamma$ spectra
from continuum and $B\Bbar$ events.}{fig:bsgamma-srcs}


\Fig{fig:bsgamma-srcs} shows that
the $\btosgamma$ signal is swamped by $\gamma$'s from continuum events,
 arising
either from $\piz$ decays or initial state radiation (ISR). Below about 2.3 GeV,
the number of $\gamma$'s from other $B\Bbar$ decays is also larger than
the $\btosgamma$ signal.   

Two basic strategies were used to reduce the huge background of 
$\gamma$'s from continuum events: combining event shapes and the energies  in
cones relative to the photon direction in a neural net (NN), and
pseudo-reconstruction (PR) -- approximately reconstructing an 
$X_s$ state from 1-4 pions and either a $\Kzs \to \pip\pim$ decay or a charged
particle with ionization consistent with a $K^\pm$.   For reconstructed events,
the $\chi^2$ derived from the pseudo-reconstruction was combined with other
kinematic variables in a neural net. If a lepton was present in either an NN
or PR event, lepton kinematic variables were also added to the appropriate
neural net.  Eventually four neural nets were used to handle the different
cases NN and PR, with and without a lepton.   Then all information from these
four neural nets was combined into a single weight between 0 (continuum) and 1
($\BtoXsgamma$).  \Fig{fig:bsgamma-w}(a) shows the distributions of these weights
versus $E_\gamma$ for On-$\Upsilon(4S)$ and Off-$\Upsilon(4S)$ (continuum) data.
\Fig{fig:bsgamma-w}(b) shows the result of subtracting the Off data from the On data
and the Monte Carlo estimate of the background from
$B\Bbar$ events.  The Subtracted and $B\Bbar$ distributions agree very well
below and above the $\btosgamma$ signal region, demonstrating that the
Continuum contribution has been estimated very accurately.  {\it Clearly the
large continuum data sample is essential for this analysis.}  

\Begfigure{t}
\includegraphics[width=\Figwidth]{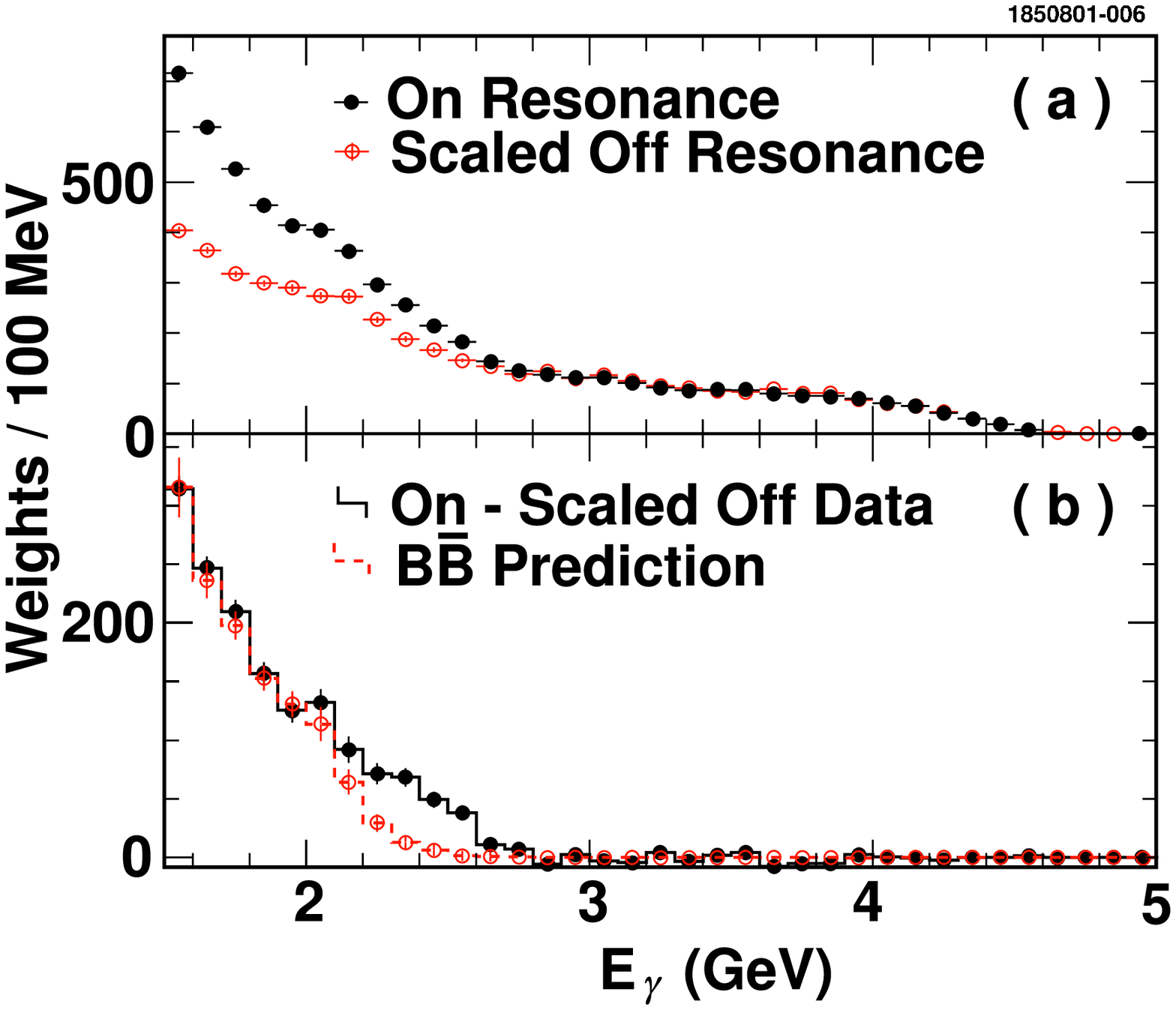}
\Endfigure{Distributions of weights versus $E_\gamma$ for
$\BtoXsgamma$ candidate events.  (a) the weight distributions for
On-$\Upsilon(4S)$ events and Off-$\Upsilon(4S)$ events scaled to the same
luminosity and CM energy.  (b) the result of subtracting the scaled
Off data from the On data and the Monte Carlo prediction of the
$B\Bbar$ contribution.\vspace*{-5ex}}{fig:bsgamma-w}

The weight distribution after subtracting the continuum and $B\Bbar$ background
contributions is illustrated in \Fig{fig:bsgamma-spect} along with the spectrum
shape derived from a Monte Carlo simulation based on the
Ali-Greub~\cite{ali-greub:bsgamma} spectator model.  Hadronization of the
$s\bar{q}$ state was modeled with $K^*$ resonances chosen to approximate the 
Ali-Greub $X_s$ mass distribution, and with JETSET tuned to the same mass
distribution.  (Very similar results are obtained from the
Kagan-Neubert~\cite{kagan-neubert} theory.)  After correcting the results for
the $b \to d\,\gamma$ contribution and the fraction of the total
$\btosgamma$ spectrum in our $E_\gamma$ interval, we obtain the branching
fraction\break

\Begfigure{t}
\includegraphics[width=\Figwidth]{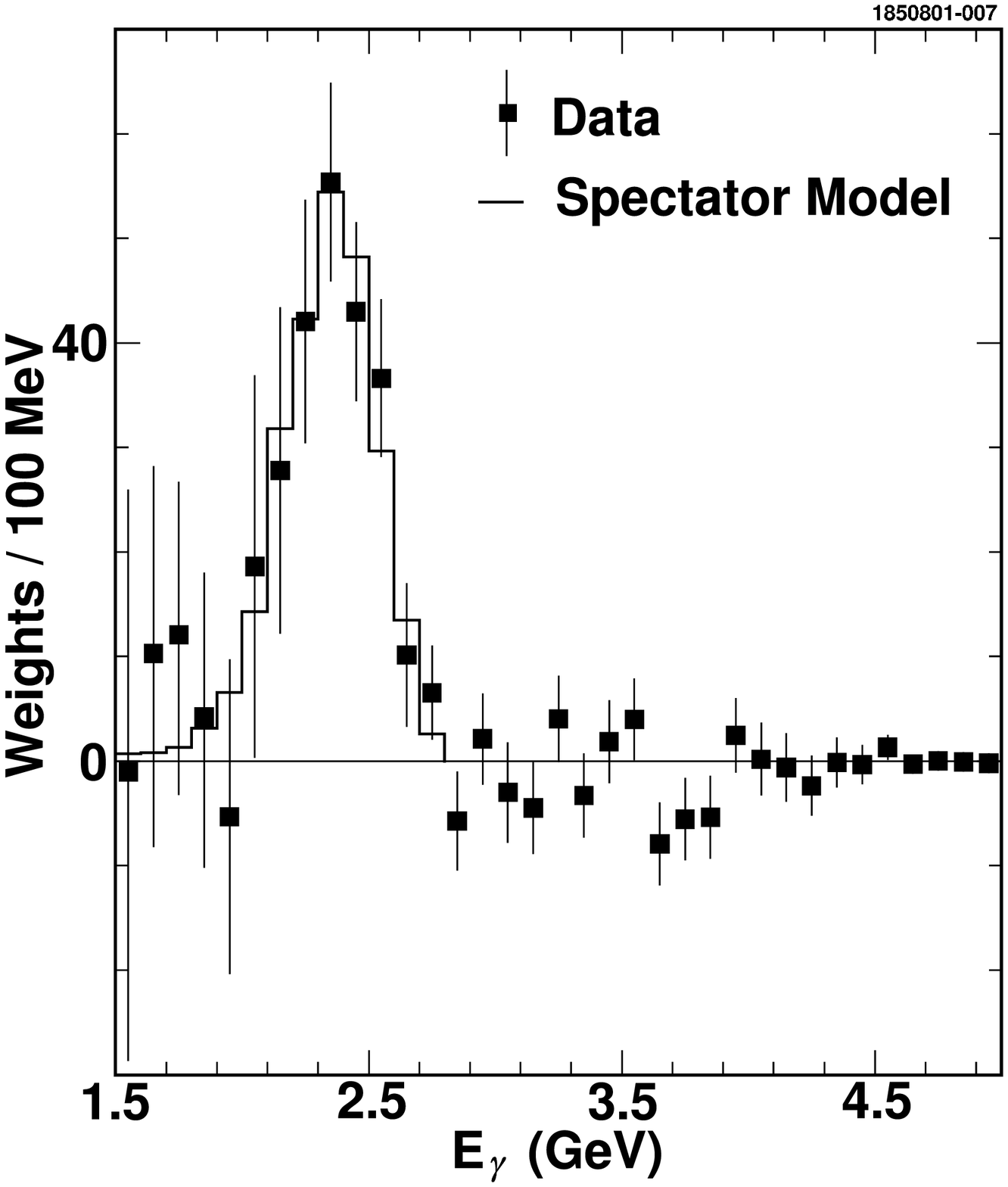}
\Endfigure{The observed $E_\gamma$ weight distribution after subtraction of
continuum and $B\Bbar$ backgrounds.  The Spectator Model spectrum
is from a Monte Carlo simulation of the
Ali-Greub~\protect\cite{ali-greub:bsgamma}
model.\vspace*{-11.5ex}}{fig:bsgamma-spect}

\Begeqnarray
\calB(\btosgamma) &=& 
(3.21 \pm 0.43 \pm 0.27 ^{+0.18}_{-0.10}) \nonumber \\ 
&\times& 10^{-4}
\Endeqnarray
where the first error is statistical, the second is systematic, and the third
is from theoretical corrections.
  
This CLEO measurement agrees very well with the previous CLEO
result~\cite{cleo:bsgamma1}. \Fig{fig:Bbsgamma} illustrates the excellent agreement of
this result with recent ALEPH~\cite{aleph:bsgamma} and Belle~\cite{belle:bsgamma}
measurements, as well as with two NLO theoretical calculations by
Chetyrkin-Misiak-M\"unz~\cite{CMM-bsgamma} (CMM) and 
Gambino-Misiak~\cite{GM-bsgamma} (GM). The agreement of this measurement with the 
theoretical predictions leaves little room for New Physics in $\btosgamma$ decays.

\vspace*{-2ex}
\section{Measuring \boldmath$\Vcb$ using Hadronic Mass Moments and $\btosgamma$}
\label{sec:Vcb-mx}

The spectator diagram for $\BtoXclnu$ decay is illustrated in \Fig{fig:bclnu-spect}.
The width $\Gamma_{SL}^c \equiv \Gamma(\BtoXclnu)$ for inclusive semileptonic
decay to all  charm states
$X_c$ is
\Begeqn
\Gamma_{SL}^c = \calB(\BtoXclnu) / \tau_B  = {\gamma_c} \Vcb^2.
\label{eq:gammacsl-gam}
\Endeqn

\twocolumn[
\vspace*{-0.9cm}
\begin{center}
\setlength{\unitlength}{0.85pt}
\newsummarypar{0}{95}{95}{0}{0}{16}                             
\begsummary     
\normalsize                                                                                                                                                             
\titline{ }{${\cal B}(b \to s \gamma)~  [10^{-4}]$}                                                                                                                                                                     
\titskip                                                                                                                                                                        
\datline{       ALEPH   \hfill  }{$     3.11     \pm    0.80     \pm    0.72    $} \fillsqrbar{ 156     }{      40      }{      54      } 
\datline{       Belle   \hfill  }{$     3.36     \pm    0.53     \pm    0.68    $} \fillsqrbar{ 168     }{      27      }{      43      } 
\divdash                                                                                                                                                                        
\datline{       CLEO II \& II.V \hfill  }{$     3.21     \pm    0.43     \pm    0.32    $} \fillcirbar{ 161     }{      22      }{      27      } 
\divline                                                                                                                                                                        
\datline{       CMM Theory               }{$    3.28     \pm    0.33                    $}    \dashbar{        148     }{      181    
}                                                               
\datline{       GM~~~ Theory             }{$    3.73     \pm    0.30                     $}    \dashbar{        172     }{      202}
\drawframe{${\cal B}~[10^{-4}]$ }                                                                                                                                                                       
        \majortick{     0       }{      0.0     }               \minortick{     25      }                                                                                               
        \majortick{     50      }{      1.0     }               \minortick{     75      }                                                                                               
        \majortick{     100     }{      2.0     }               \minortick{     125     }                                                                                               
        \majortick{     150     }{      3.0     }               \minortick{     175     }                                                                                               
        \majortick{     200     }{      4.0     }               \minortick{     225     }                                                                                               
        \majortick{     250     }{      5.0     }                                                                                                                               
\endsummary
\bigskip
\Figurecaption{Comparison of the new CLEO measurement of $\calB(\btosgamma)$ to
the ALEPH~\protect\cite{aleph:bsgamma} and Belle~\protect\cite{belle:bsgamma}
measurements and to the CMM~\protect\cite{CMM-bsgamma} and
GM~\protect\cite{GM-bsgamma} Standard Model theory calculations.}\label{fig:Bbsgamma}
\end{center}
\bigskip]

\Begfigure{htb}
\imespect{$\bar{B}$}
\iqespect{$b$}{$\bar{q}$}\wespect
\raisebox{0.3ex}{\includegraphics[width=10em]{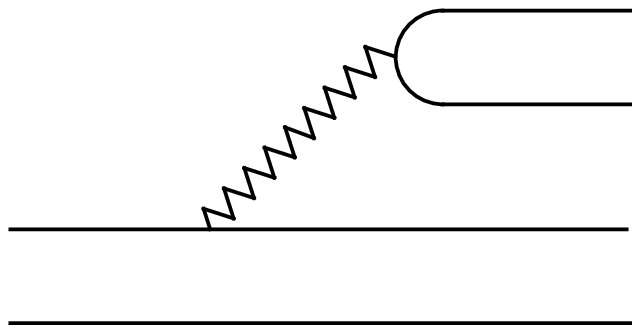}}
\fqespect{$\bar{\nu}_e$}{$e$}{$c$}{$\bar{q}$}
\fqespect{~$\bar{\nu}_\mu$}{~$\mu$}{}{}
\fmespect{}{$X_c$}
\Endfigure{The spectator diagram for $\BtoXclnu$ decay.  The same diagram
with $c$ replaced with $u$ describes $\BtoXulnu$
decay.\vspace*{-2ex}}{fig:bclnu-spect}

Clearly the CKM matrix element $\Vcb$ can
 be determined from
the branching fraction for $\BtoXclnu$ decays, if the theoretical parameter
$\gamma_c$ is known.  Unfortunately
$\gamma_c$ is a nonperturbative QCD parameter, and theoretical
models have been the only means of estimating $\gamma_c$.  However, measurements
of hadronic mass moments in
$\BtoXclnu$ decays combined with energy moments in   
$\btosgamma$ decays can essentially eliminate model dependence.

To order $1/M_B^3$ the decay width
$\Gamma(\Bbar \to X_c \ell \nubar)$ can be written in the form
\Begeqnarray
\Gamma_{SL}^c 
&=& {G_F^2 \Vcb^2 M_B^5 \over 192 \pi^3} 
\left[\calG_0 
+ {1 \over M_B}\calG_1(\Lambdabar)\right. \nonumber\\ 
&+& {1 \over M_B^2}\calG_2(\Lambdabar,\lambda_1,\lambda_2) \label{eq:gammacsl-lam}\\ 
&+& \left. {1 \over M_B^3}\calG_3
(\Lambdabar,\lambda_1,\lambda_2|
\rho_1,\rho_2,\calT_1,\calT_2,\calT_3,\calT_4)\right] \nonumber 
\Endeqnarray
where
$\Lambdabar,\lambda_1,\lambda_2,
\rho_1,\rho_2,\calT_1,\calT_2,\calT_3,\calT_4$ are nonperturbative
QCD parameters,
the $\calG_n$ are polynomials of order $\leq n$ in $\Lambdabar,\lambda_1,\lambda_2$, 
and $\calG_3$ is linear in $\rho_1,\rho_2,\calT_1,\calT_2,\calT_3,\calT_4$.
Some of the coefficients of the polynomials $\calG_n$ involve expansions in
$\alpha_S$.

There are similar expressions -- involving the same nonperturbative QCD parameters
-- for the moments
$\langle (M_X^2 - \bar{M}_D^2) \rangle$ of the hadronic mass ($M_X$)
spectrum in $\BtoXclnu$ decay and 
$\langle E_\gamma \rangle$ of the energy spectrum in  $\btosgamma$ decay.  (Here
$\bar{M}_D = 0.25 M_D + 0.75 M_{\Dstar}$, the spin-averaged $D$ meson mass.) The
coefficients $\calM_n$ and $\calE_n$ of the polynomials for these moments depend on
the lepton momentum range measured in $\BtoXclnu$ decays and the energy range
measured in
$\btosgamma$ decays, respectively.  

To obtain $\Vcb$ from \Eqn{eq:gammacsl-lam},
we determined $\Lambdabar$ and $\lambda_1$ from
$\langle (M_X^2 - \bar{M}_D^2) \rangle$ and $\langle E_\gamma \rangle$
after: determining $\lambda_2$ from $M_{\Bstar} - M_{B}$ and
estimating
$\rho_1,\rho_2,\calT_1,\calT_2,\calT_3,\calT_4$ to be about $(0.5~{\rm GeV})^3$ from
dimensional considerations.

Moments of the $E_\gamma$ spectrum in $\btosgamma$ decay were determined
from the data and the spectator model illustrated in 
\Fig{fig:bsgamma-spect} in the previous section.  The first and
second moments of $E_\gamma$ obtained in this analysis are:
\Begeqnarray
\langle E_\gamma \rangle~~~~~~~ &=& 2.346\;\, \pm 0.032\;\, \pm 0.011~~\, 
\nonumber \\
&& {\rm GeV~~and} \\
\langle (E_\gamma -\langle E_\gamma \rangle)^2 \rangle &=& 
0.0226 \pm 0.0066 \pm 0.0020\;~~ \nonumber \\
&& {\rm GeV}^2. 
\Endeqnarray

The calculation of the hadronic mass $M_X$ starts with reconstruction of the 
neutrino in events with a single lepton by ascribing the missing energy and momentum
to the neutrino.  We then use 
$M_X^2 \cong M_B^2 + M_{\ell\nu}^2 - 2 E_B E_{\ell\nu}$ where $M_{\ell\nu}$ and
$E_{\ell\nu}$ are the invariant mass and the energy of the $\ell\nu$ system,
respectively.  (This expression is obtained by setting 
$\cos\theta_{B-\ell\nu} = 0$, where $\theta_{B-\ell\nu}$ is the unmeasurable 
angle between the momenta of the $B$ and the $\ell\nu$ system.)  
Neutrino energy and momentum resolution, and neglect of the modest term involving
$\cos\theta_{B-\ell\nu}$ result in non-negligible width for the $M_X$ distributions
of $\BtoDlnu$ and $\BtoDstarlnu$ decays.

\Begfigure{htb}
\includegraphics[width=\Figwidth]{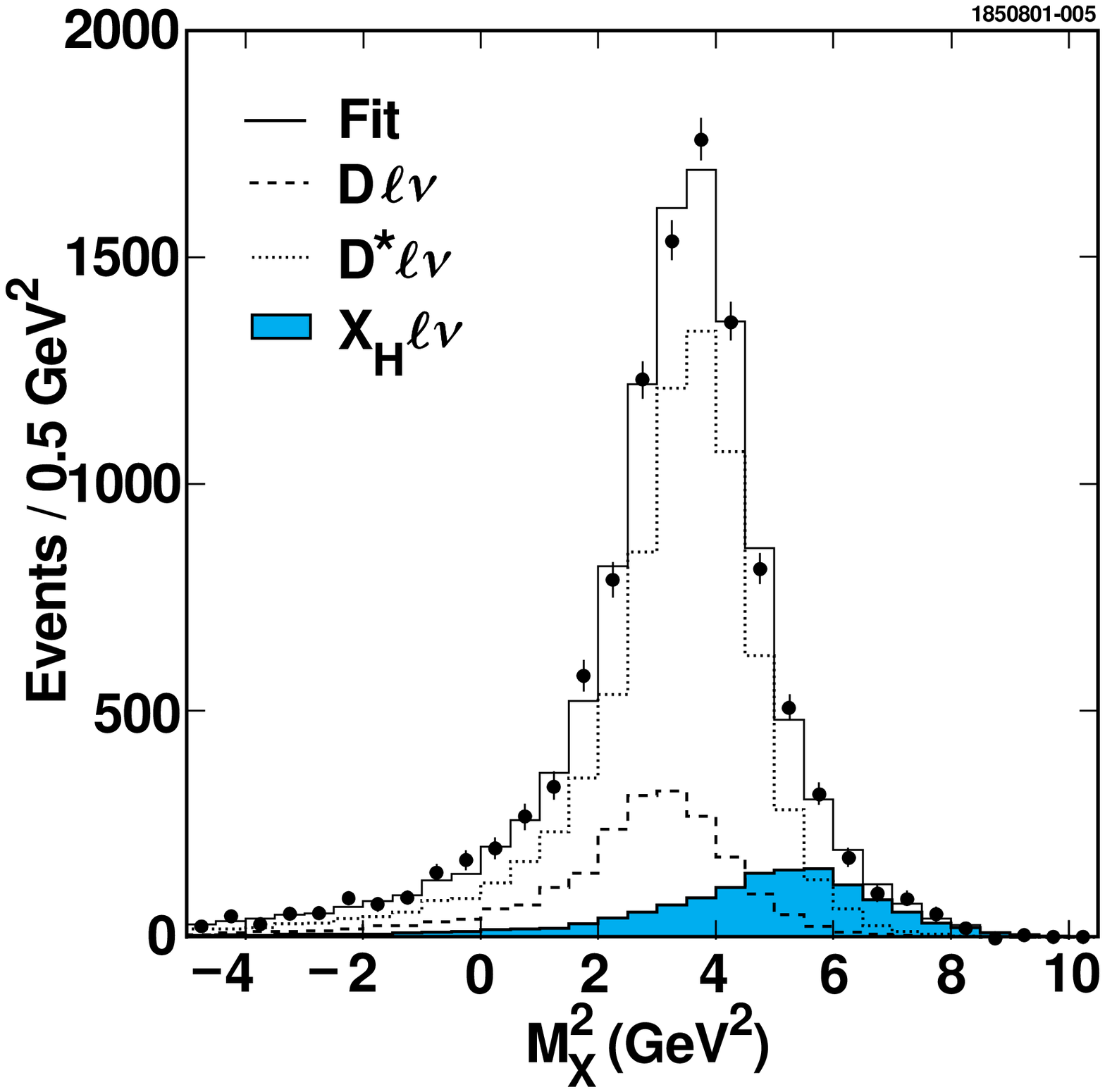}
\Endfigure{The measured $M_X^2$ distribution (points), Monte Carlo
simulation (solid line), and the three components of the Monte Carlo
simulation -- $\Bbar \to D \ell \nu$ (dashed), 
$\Bbar \to \Dstar \ell \nu$ (dotted) and
$\Bbar \to X_H \ell \nu$ (shaded).}{fig:mxsq}

\vspace*{-2ex}

\Fig{fig:mxsq} illustrates the experimental\cite{cleo:mxsq} $M_X^2$ distribution
and Monte Carlo simulation using contributions from 
$\Bbar \to D \ell \nu$,
$\Bbar \to \Dstar \ell \nu$, and
$\Bbar \to X_H \ell \nu$ decays, where $X_H$ denotes high mass resonant and
nonresonant charm meson states with masses above the $\Dstar$.  
The relative amounts of the $D$, $\Dstar$, and $X_H$ contributions are determined
in fits to the data.  The relative rates and the generated masses are used
to calculate the hadronic mass moments. The
relative rates are sensitive to the model used for the 
$\Bbar \to X_H \ell \nu$ spectrum, but the
$M_X^2$ moments are quite insensitive.  The dispersions in the moments for
different $\Bbar \to X_H \ell \nu$ models are included in the systematic errors for
the moments.   The first and second moments of $M_X^2$ obtained from this analysis
are
\Begeqnarray
\langle (M_X^2 - \bar{M}_D^2) \rangle\;\,  &=& 0.251 \pm 0.023 \pm 0.062~~~~~~~
\nonumber \\ 
&& {\rm GeV^2~~and} \\
\langle (M_X^2 - \bar{M}_D^2)^2 \rangle  &=& 
0.639 \pm 0.056 \pm 0.178\, \nonumber \\
&& {\rm GeV}^4 
\Endeqnarray

The experimental moments were measured with $E_\ell > 1.5$ GeV and
$E_\gamma > 2.0$ GeV.  Falk and Luke~\cite{falk-luke} calculated the
coefficients of the polynomials $\calE_n$ and $\calM_n$ for the same ranges of
$E_\ell$ and $E_\gamma$.  
We use only $\langle E_\gamma \rangle$ and $\langle (M_X^2 - \bar{M}_D^2) \rangle$
to determine $\Lambdabar$ and $\lambda_1$ since theoretical expressions for the
higher moments converge slowly and are much less reliable~\cite{falk-luke}. 
These moments define the bands in the $\Lambdabar$-$\lambda_1$ plane, illustrated in
\Fig{fig:lambda-bands}.  The intersection of the bands from the two moments
yields
\Begeqnarray
\Lambdabar~ &=&~~ 0.35~\; \pm 0.07~\, \pm 0.10~~~\,  {\rm GeV}~~~~~~~~~~~ \\
\lambda_1 &=& -0.236 \pm 0.071 \pm 0.078~~  {\rm GeV^2}
\Endeqnarray
where the errors are experimental and theoretical in that order.

The other experimental measurements we used to determine $\Vcb$ are: 
\Begeqn
\calB(\BtoXclnu) = (10.39 \pm 0.46)\%
\Endeqn
from CLEO~\cite{cleo:B-bclnu}; the ratio
\Begeqn
(f_{+-} \tau_{\Bm})/(f_{00} \tau_{\Bz}) = 1.11 \pm 0.08
\Endeqn
from CLEO~\cite{cleo:ftauratio}, where
\Begeqnarray
f_{+-} &\equiv& \calB(\Upsilon(4S)\! \to\! \Bp\Bm)~~ \mathrm{and} \\
f_{00}\hspace*{0.4em} &\equiv& \calB(\Upsilon(4S)\! \to\! \Bz\Bzbar);
\Endeqnarray
and the PDG~\cite{PDG} average values of
$\tau_{\Bm}$ and $\tau_{\Bz}$. From this analysis we obtain
\Begeqn
\Vcb = (40.4 \pm 0.9 \pm 0.5 \pm 0.8) \times 10^{-3} \label{eq:Vcb-inc}
\Endeqn
where the errors are due to uncertainties in moments, $\Gamma^c_{SL}$, and theory
(the $\alpha_s$ scale and\break
\Begfigure{tb}
\includegraphics[width=\Figwidth]{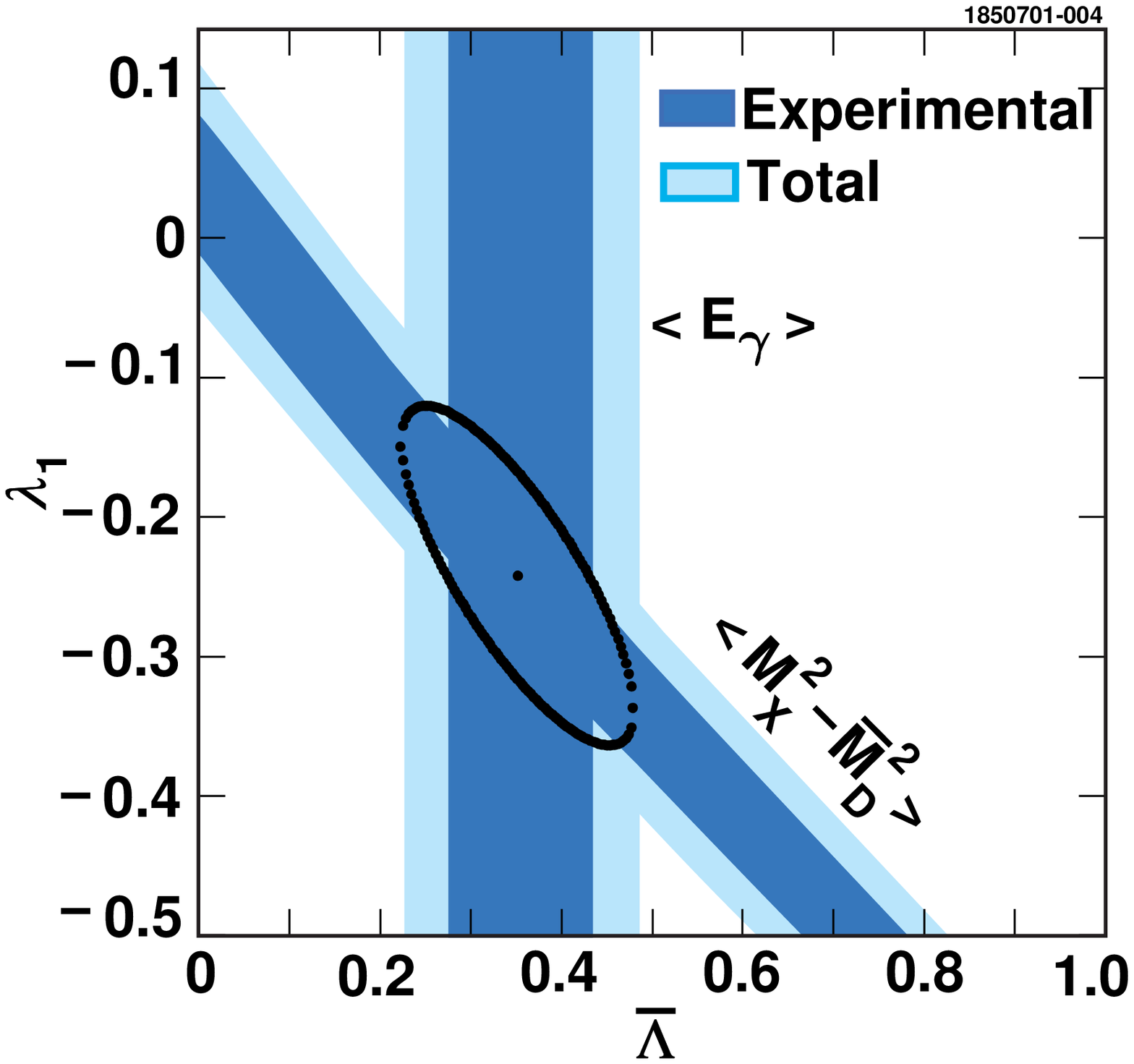}
\Endfigure{Bands in $\Lambdabar$ and $\lambda_1$ defined by the measured
$\langle E_\gamma \rangle$ and $\langle (M_X^2 - \bar{M}_D^2) \rangle$ moments.
The dark gray bands indicate the experimental errors and the light gray
extensions illustrate the contributions of the theoretical
uncertainties.\vspace*{-5ex}}{fig:lambda-bands}
\noindent $\calO(1/M_B^3)$ terms) in that order.
This result agrees well with earlier measurements based on models, indicating
that the models are reasonably adequate.

In the question period following this report, M.\ Wise pointed out that
moments of the lepton momentum spectrum depend on $\Lambdabar$
and $\lambda_1$ and asked if we were also using these moments. 
Measurement of these moments is quite sensitive to systematic errors and we are
working to control these errors.\\[-5.5ex]

\section{\boldmath$\Vub$ from Inclusive Leptons and the $\btosgamma$ Spectrum}
\label{sec:Vub}

Simply by replacing $c$ with $u$, the spectator diagram (\Fig{fig:bclnu-spect})
and the expression for the semileptonic width (\Eqn{eq:gammacsl-gam}) that
describe $\BtoXclnu$ decays also describe $\BtoXulnu$ decays. The CKM matrix
element $\Vub$ can then be determined from $\calB(\BtoXulnu)$
and $\gamma_u$.  However, measuring $\Vub$ is much more difficult because the rate
of $\BtoXulnu$ decays is only about 1\% of the $\BtoXclnu$ rate.  Two methods have
been used to measure $\Vub$: measuring the inclusive lepton momentum ($p_\ell$)
spectrum above or near the $\BtoXclnu$ endpoint or studying
exclusive $\Bbar \to \pi(\rho) \ell \nubar$ decays.  So far it has not been possible
to separate exclusive decays with low $p_\ell$ from background, so
either way theory is required for the fraction $f_u(p)$ of the $p_\ell$ spectrum
that lies in an interval $(p)$ above some cut.
Theoretical models for $f_u(p)$ have large uncertainties that are difficult to
quantify, leading to  severe model dependence in determining $\Vub$.
However, it is possible to eliminate most of this
uncertainty for inclusive $\BtoXulnu$ decays using $\btosgamma$ decays.  
The shape function that relates parton-level $\btosgamma$ decays to observed
$\BtoXsgamma$ decays also relates parton-level $\btoulnu$ decays to 
$\BtoXulnu$ decays.  The strategy for determining $\Vub$ is then to
fit the $E_\gamma$ spectrum (\Fig{fig:bsgamma-spect}) from the $\BtoXsgamma$
analysis in \Sec{sec:bsgamma} to a shape function~\cite{kagan-neubert} and then to
use the shape parameters to determine $f_u(p)$~\cite{fazio-neubert}.
Then the $\BtoXulnu$ branching fraction $\calB_{ub} \equiv \calB(\BtoXulnu)$ can be
determined from the measured branching fraction $\Delta\calB_{ub}(p)$ for
$\BtoXulnu$ in the momentum interval $(p)$, using
$\calB_{ub} =\Delta\calB_{ub}(p)/f_u(p)$.

\Fig{fig:bulnu-endpoint} illustrates the lepton momentum spectra in the region
above 2.0 GeV/$c$ from the full \Cleoii\ and II.V data samples.  Above about
2.3  GeV/$c$ the background is dominated by leptons and fake leptons from
Off-$\Upsilon(4S)$ (continuum) events.  At lower momenta leptons from $\BtoXclnu$
and fake leptons from hadronic $B\Bbar$ decays dominate the background.  
We use the momentum interval 2.2 - 2.6 GeV/$c$ and determine the 
{\it preliminary} partial branching fraction 
$\Delta\calB_{ub}(p) = (2.35 \pm 0.15 \pm 0.45) \times 10^{-4}$,
where the errors are statistical and systematic in that order.  From fits to the
$\BtoXsgamma$ spectrum we obtain $f_u(p) = 0.138 \pm 0.034$, where the
error includes combined experimental and theoretical uncertainties.  The value of
$\calB_{ub}$ derived from these numbers is also divided by a factor of $(0.95 \pm
0.02)$ to correct for QED radiative corrections.  To obtain $\Vub$ from
$\calB_{ub}$ we use
\Begeqnarray
\Vub &=& \left[(3.06 \pm 0.08 \pm 0.08) \times 10^{-3} \right]~~~~~~~~~
\nonumber\\
&\times& \left[\frac{\calB_{ub}}{0.001}~
\frac{1.6 {\rm~ps}}{\tau_B}\right]^\frac{1}{2} \label{eq:vcb}
\Endeqnarray
from Hoang, Ligeti, and Manohar~\cite{HLM}.
(Uraltsev~\cite{uraltsev} obtained a nearly identical result.) The {\it
preliminary} result is
\Begeqn
V_{ub} = (4.09 \pm 0.14 \pm 0.66) \times\! 10^{-3}
\Endeqn
where the first error is statistical and the second is systematic.  This result is
in good agreement with previous CLEO measurements~\cite{cleo:Vub-1993} based on
theoretical models. Currently we are working to extend the momentum range to include
as much of the
$\BtoXulnu$ spectrum as possible, in order to minimize the residual theoretical
uncertainties.  

\Begfigure{t}
\includegraphics[width=\Figwidth]{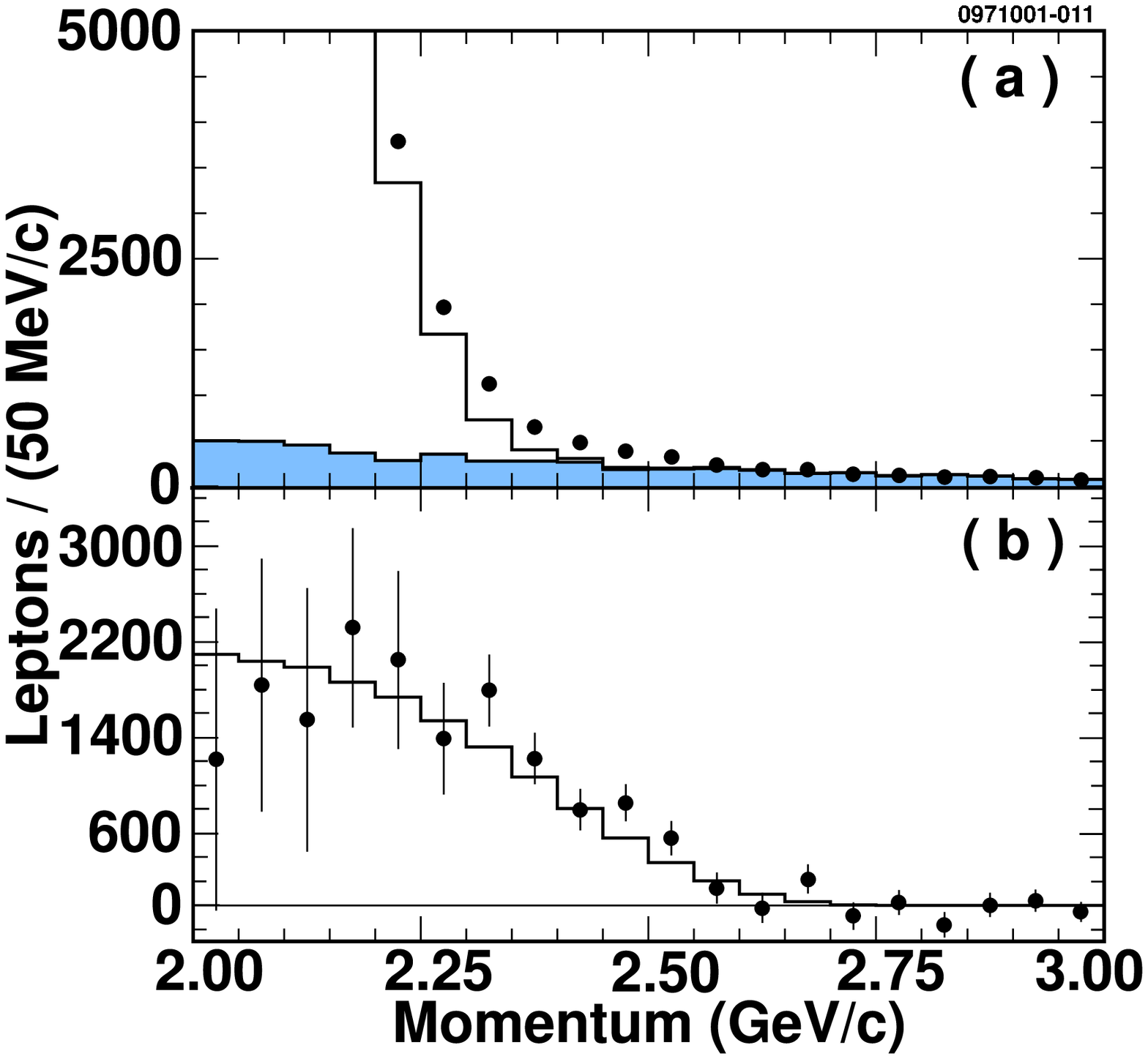}
\Endfigure{(a) The lepton momentum spectra for On-$\Upsilon(4S)$ data
(filled circles), scaled Off-$\Upsilon(4S)$ data (shaded
histogram), and sum of scaled-Off and backgrounds from $B$ decays (solid histogram).
(b) The lepton spectrum from On-$\Upsilon(4S)$ data after subtracting
Off-$\Upsilon(4S)$ and $B\Bbar$ backgrounds and correcting for efficiency (filled
points) and the $\BtoXulnu$ spectrum derived from the $\BtoXsgamma$ spectrum
(histogram).\vspace*{-4ex}}{fig:bulnu-endpoint}

\section{\boldmath$\Vcb$ from $\BtoDstarlnu$ Decay}

Exclusive $\BtoDstarlnu$ decays provide another method for measuring $\Vcb$ with
experimental and theoretical uncertainties that are substantially different from
those in inclusive measurements.  The key to these measurements is utilization
of the Isgur-Wise symmetry and Heavy Quark Effective Theory
(HQET)~\cite{neubert:phys-rep}. From HQET, the differential decay width
for $\BtoDstarlnu$ decay is
\Begeqn
{d \Gamma(w) \over dw} = 
{G_F^2 \over 48 \pi^3}\; \calG(w)\;  {\Vcb^2}\;
{\FDs^2(w)}  \label{eq:vcbfw}
\Endeqn
where $w \equiv v_B \cdot v_{\Dstar}$ 
($v_B$ and $v_{\Dstar}$ are the four-velocities of the $B$ and $\Dstar$),
$\calG(w)$ is a known function of $w$, and
$\FDs(w)$ is a nonperturbative QCD form factor that
parameterizes the $w$ dependence of the hadronic current.  The variable
$w$ is related to more familiar variables in  $\BtoDstarlnu$ decay via 
\Begeqn
w  =  {E_{\Dstar}
\over M_{\Dstar}}  = { M_B^2 + M_{\Dstar}^2 -q^2 \over 2 M_B M_{\Dstar}}
\Endeqn
where $E_{\Dstar}$ is the energy of the $\Dstar$ in the $B$ rest frame
and $q^2$ is the momentum transfer, \ie, the square of the mass of the
$\ell\nubar$ system.  The range of $w$ is $(1.00 < w < 1.504)$ for $\BtoDstarlnu$
decay.

Without HQET, three unknown form factors appear in the  differential decay width for
$\BtoDstarlnu$.  
These three form factors are related by HQET, resulting in the 
the simple form given in \Eqn{eq:vcbfw} with only one unknown function 
$\FDs(w)$.  Furthermore, for large heavy quark
masses $m_Q$, $\FDs(w)$ is constrained by HQET at
$w = 1$ (or $q_{\rm max}^2$) to
\Begeqn
\FDs(1) \approx \eta_A [ 1 + \delta(1/m_Q^2) ]
\Endeqn
where $\eta_A$ is a perturbative QCD correction, and $\delta(1/m_Q^2)$ is a
nonperturbative QCD correction of $\calO(1/m_Q^2)$.  

Hence, the ideal strategy for determining $\Vcb$ using HQET would be to
measure $d\Gamma(w)/dw$ at $w=1$. However, $\calG(1) = 0$ due to phase space,
so the practical strategy is to measure $d\Gamma(w)/dw$,
determine $\Vcb\FDs(w)$ from a fit over the full $w$ range, and extrapolate 
it to $w=1$ to obtain $\Vcb\FDs(1)$.
Following Caprini-Lellouch-Neubert~\cite{CLN} (CLN), the dependence of $\FDs(w)$ on
$w$ can be reduced to dependence on two ratios of form factors $R_1(1)$ and
$R_2(1)$, previously measured by CLEO~\cite{cleo:r1r2}, and a slope parameter
$\rho^2$ to be determined in the fit. 

The branching fraction for $\BtoDstarlnu$ decay and the product $\Vcb\FDs(w)$
has been measured  before~\cite{PDG}.  We now measure these quantities with
a larger data sample (3.0 \fbinv\ of \Cleoii\ data) and substantially reduced
systematic errors.  In addition, we now report 
results for both $\BztoDstarplnu$ and $\BmtoDstarzlnu$ decays.  
Previously~\cite{cleo:ichep2000}, we reported results for $\BztoDstarplnu$
based on this full data sample.

We reconstruct $\BtoDstarlnu$ decays by
finding $\Dstar$ candidates using $\Dstar \to \Dz \pi$ and 
$\Dz \to \Km \pip$ decays,
and finding an $e$ or a $\mu$ in the event with the same sign as the $\Km$ and
$p_e > 0.8$ GeV/$c$ or $p_\mu > 1.4$ GeV/$c$. We
separate the $\BtoDstarlnu$ signal from background using the angle 
$\theta_{B-\Dstar\ell}$ between the momenta of the
$B$ and the $\Dstar\ell$ combination.
The cosine of $\theta_{B-\Dstar\ell}$ is 
\Begeqn
\cos\theta_{B-\Dstar\ell} = 
{2 E_B E_{\Dstar\ell} - M_B^2 - M_{\Dstar\ell}^2 \over 
2\, P_B\; P_{\Dstar\ell}} 
\Endeqn
where $E_B$, $P_B$, $M_B$, $E_{\Dstar\ell}$, $P_{\Dstar\ell}$, and 
$M_{\Dstar\ell}$ are the energy, momentum, and mass of the $B$ and the
$\Dstar\ell$ system, respectively.
\Fig{fig:cosBD*l} shows the $\cos\theta_{B-\Dstar\ell}$ distributions
for the $w$ range, $1.10 < w < 1.15$.  We fit the data with distributions
for $\Dstar\ell\nu$ signals and 5 different types of backgrounds.  The
background shapes are determined from a combination of Monte Carlo calculations and 
background data samples.  The combinations of signals and backgrounds fit the
data very well in all $w$ intervals.

The values of $\Vcb\FDs(w)$ obtained from the fits to the
$\cos\theta_{B-\Dstar\ell}$ distributions are illustrated in \Fig{fig:VcbFw}
along with the fit to the  $\Vcb\FDs(w)$ data. 
The ingredients used in the fit are the shape of
$\FDs(w)$ from Caprini-Lellouch-Neubert~\cite{CLN},
$\FDs(w) = \calF_{\Dstarp}(w) = \calF_{\Dstarz}(w)$,
$\Gamma(\Dstar\ell\nubar) = \Gamma(\Dstarp\ell\nubar) = \Gamma(\Dstarz\ell\nubar)$,
and the CLEO measurement~\cite{cleo:ftauratio} of the ratio
$(f_{+-} \tau_{\Bm}) / (f_{00} \tau_{\Bz})$.

From the fit to the $\Vcb\FDs(w)$ distribution, we obtain the {\it preliminary}
results:
\Begeqnarray
\Vcb\FDs(1) &=& (42.2 \pm 1.3 \pm 1.8) \times 10^{-3},~~~~~~~  \\
\rho^2 &=& \;\,1.61 \pm 0.09 \pm 0.21,~         \textrm{and} \\
\Gamma(\Dstar\ell\nubar) &=& \nonumber\\
(0.037.6 &\pm& 0.001.2 \pm 0.002.4)~ {\rm ps}^{-1},
\Endeqnarray

\Begfigure{t}
\includegraphics[width=\Figwidth]{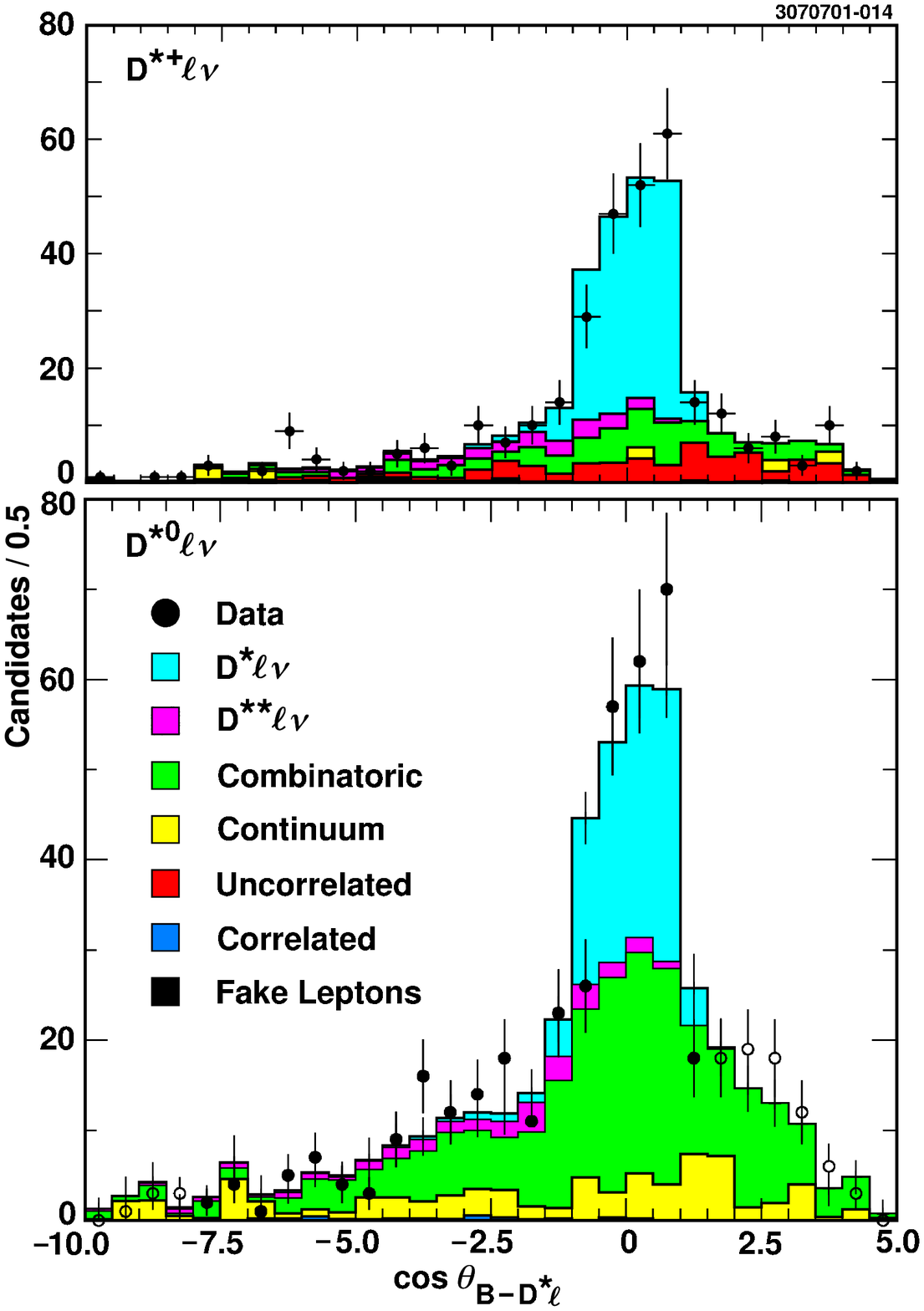}
\Endfigure{Distributions of $\cos\theta_{B-\Dstar\ell}$ for
$\BztoDstarplnu$ (top) and $\BmtoDstarzlnu$ (bottom) for the $w$ range, 
$1.10 < w < 1.15$. The filled circles are the data and the shaded histograms are the
contributions of the $\Dstar\ell\nu$ signals and the backgrounds.
\vspace*{-4ex}}{fig:cosBD*l}

\noindent where $\rho^2$ is the slope parameter for the CLN form factor,
$\FDs(w)$.
Using the PDG~\cite{PDG} average lifetimes we also obtain
{\it preliminary} branching fractions from the measured decay width,
$\Gamma(\Dstar\ell\nubar)$: 
\Begeqnarray
\calB(\Dstarz\ell\nubar) &=& (6.21 \pm 0.20 \pm 0.40)\% \\
\calB(\Dstarp\ell\nubar) &=& (5.82 \pm 0.19 \pm 0.37)\%~~~~~~~~~  
\Endeqnarray
These branching fractions and the value of $\Vcb\FDs(1)$ that we obtain
are somewhat higher than and are marginally consistent with previous measurements
from LEP~\cite{lep:hfwg}.  

The parameters $\Vcb\FDs(1)$ and $\rho^2$ are
generally highly correlated in the fits, so it is necessary to take these
correlations into account in comparing results from different experiments.
This is illustrated in
\Fig{fig:Vcbellipses}.  

\Begfigure{htb}
\includegraphics[height=6.6cm]{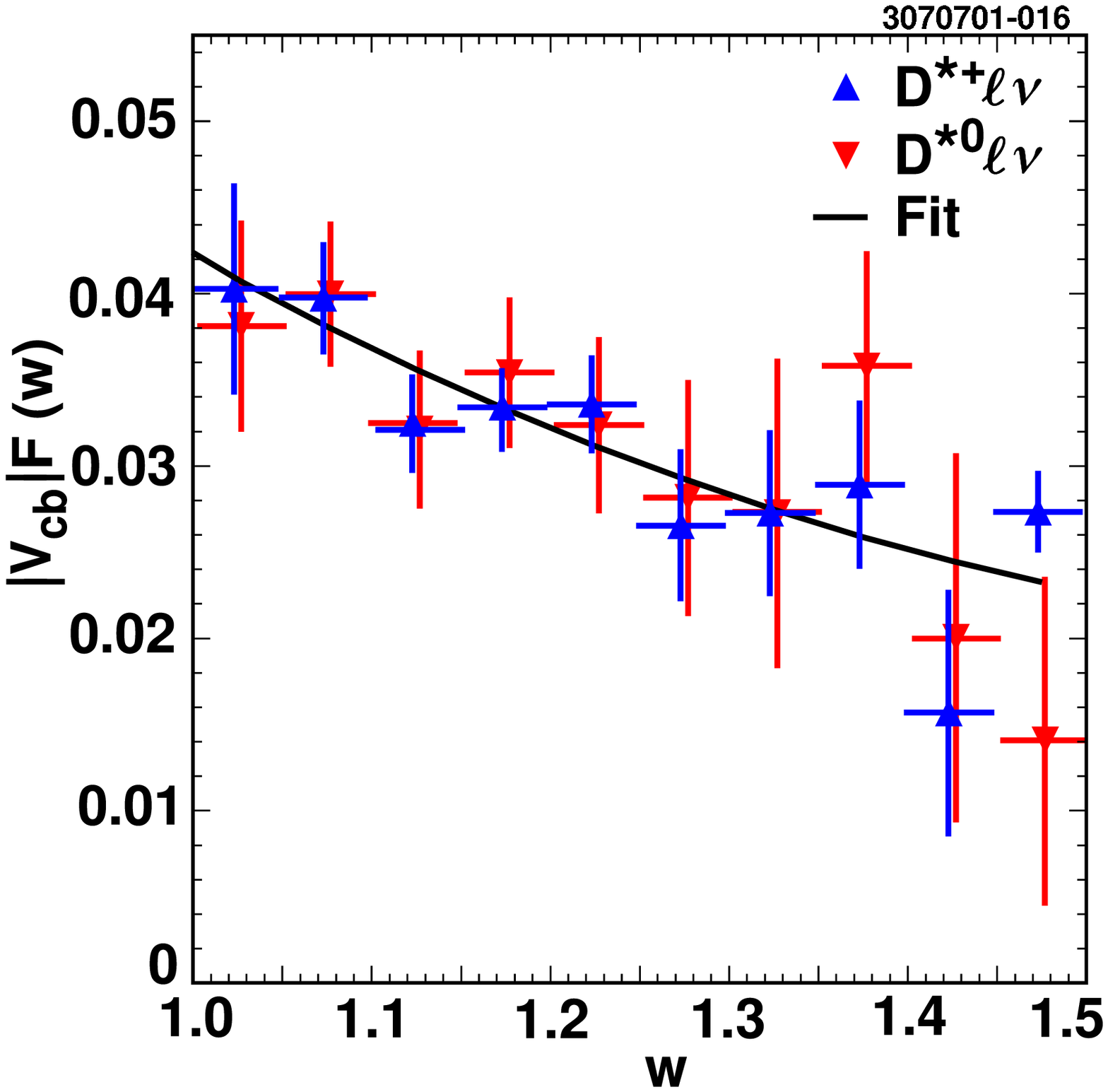}
\Endfigure{Values of  $\Vcb\FDs(w)$ obtained from the fits to the
$\cos\theta_{B-\Dstar\ell}$ distributions for $\BztoDstarplnu$ (upward triangles) and
$\BmtoDstarzlnu$ (downward triangles).  The fit is described in the 
text.\vspace*{-6ex}}{fig:VcbFw}

The correlation between
$\Vcb\FDs(1)$ and $\rho^2$ in CLEO data is less than the correlation in LEP
data. This is due to an interaction in the systematic error between the lepton
momentum cuts that we use and the measured form factor ratios
$R_1(1)$ and $R_2(1)$ in the CLN form of $\FDs(w)$. 

One possible source of the apparent discrepancy between CLEO and LEP measurements
is the fact that $\Dstar X \ell\nubar$ components are estimated differently
by the two different groups; CLEO included this component in the 
$\cos\theta_{B-\Dstar\ell}$ fit, while the LEP collaborations use a model
constrained by LEP measurements of $\Bbar \to \Dstar X \ell \nu$.  

In order to derive $\Vcb$ from the measured value of $\Vcb\FDs(1)$, we use
\Begeqn
\FDs(1)=0.913 \pm 0.042
\Endeqn
from the {\sc BaBar} Physics Book~\cite{babar:PB}
and obtain
\Begeqn
\Vcb = (46.2 \pm 1.4 \pm 2.0 \pm 2.1) \times 10^{-3}
\Endeqn
where the errors are statistical, systematic, and due to the uncertainty
in $\FDs(1)$.  The difference between this result and the inclusive measurement
(\Eqn{eq:Vcb-inc}) may indicate a breakdown of quark-hadron duality in
inclusive semileptonic $B$ decays.

\Begfigure{t}
\includegraphics[height=6.6cm]{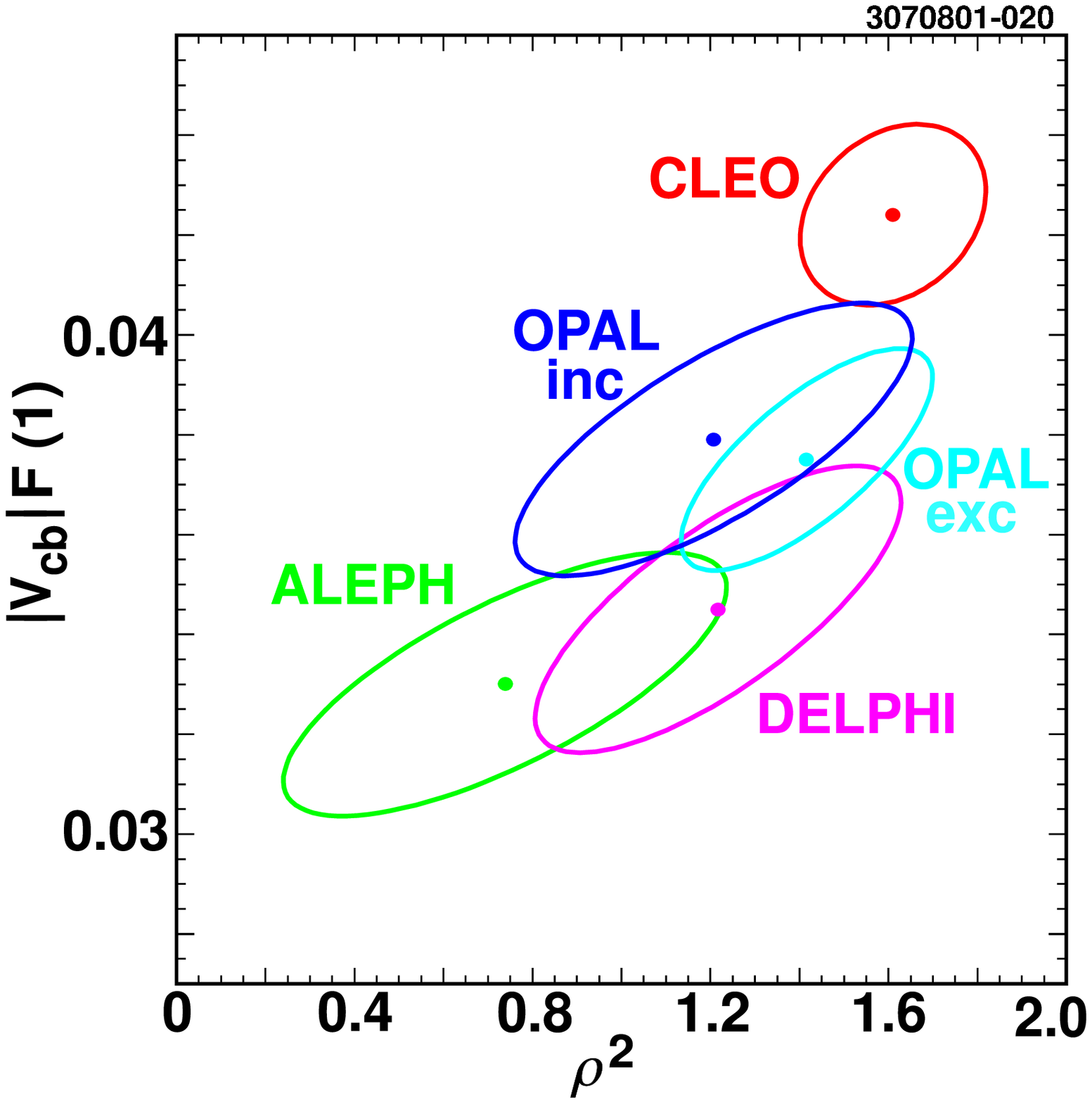}
\Endfigure{The correlations between $\Vcb\FDs(1)$ and $\rho^2$
in CLEO and LEP~\protect\cite{lep:hfwg} measurements. OPAL used a partial
reconstruction of $\BztoDstarplnu$ decays for the measurement labeled OPAL
inc.\vspace*{-2ex}}{fig:Vcbellipses}

\section{\boldmath Preliminary \Cleoiii\ Results on Rare $B$ Decays}
\label{sec:brare}

Previously CLEO measured the branching fractions for the four possible
$B \to K\pi$ modes and $B \to \pip\pim$, and determined comparable upper
limits for the other $B \to \pi\pi$ modes and for $B \to K\Kbar$ 
decays~\cite{cleo:Brare98,cleo:Brare00,cleo:Bpizpiz}.   
These exclusive two-body $B$ decays are very important because:
\Begitem
\item   Certain ratios of $B \to K\pi$ branching fractions 
depend~\cite{ref:brare,BBNS}
explicitly on the angle $\gamma = \arg(V_{ub}^*)$ of the unitarity triangle 
(\Fig{fig:unitriangle}).  The modest dependence of these ratios on models suggests
that $\gamma$ can be obtained from fits to comprehensive
measurements of these branching fractions.  
\item The sum of the angles $\beta$ and $\gamma$ (\Fig{fig:unitriangle}) can also be
determined from time-dependent $CP$ violation measurements in $\Bz \to \pip\pim$
decays.  This requires separation of penguin contributions from tree contributions
to the decay using isospin analysis~\cite{gronau-london} of all three $B \to \pi\pi$
charge states.
\item Whether or not the $CP$ violating phase in the CKM matrix is the sole source of
$CP$ violation is still an open question.  Self tagging rare decay modes such as
$\Bp \to \Kp\piz$ are an obvious arena in which to search for other manifestations of
$CP$ violation.
\Enditem
Hence, these rare $B$ decays can play particularly important
roles in constraining the CKM matrix and developing our understanding of $CP$
violation.

Only \Cleoii\ and \Cleoiiv\ data were used in previous CLEO measurements of
these $B$ decays.  CLEO now has {\it preliminary} measurements of branching
fractions or determinations of upper limits for 
$\BtoKpi$, $\Btopipi$, $B \to K\Kbar$, and 
$B^- \to \Dz\Km$ decays, from about one-half of the \Cleoiii\ data.  This is
the first public presentation of these results.
 
\Begfigure{htb}
\includegraphics[width=\Figwidth]{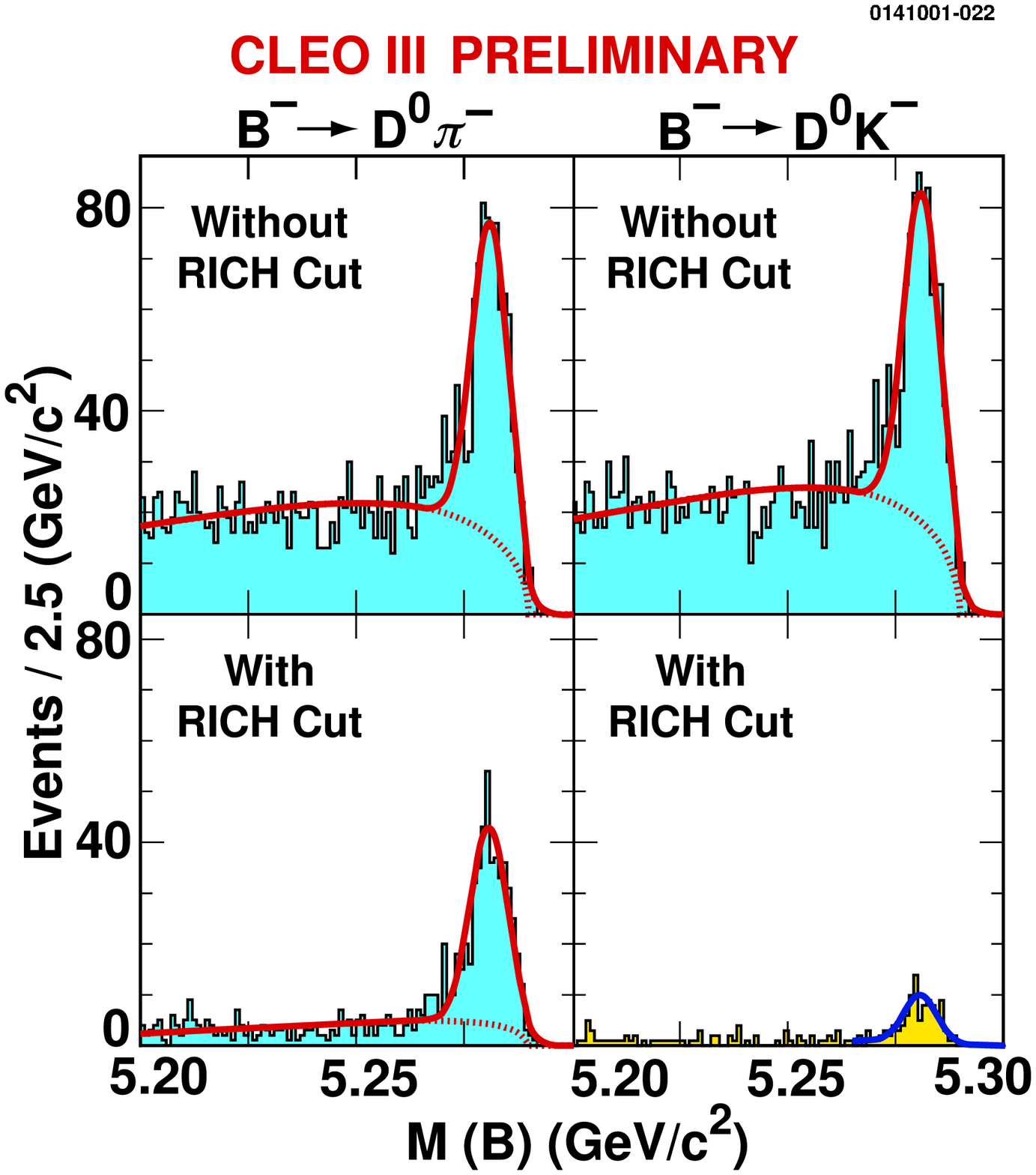}
\Endfigure{Preliminary \Cleoiii\ mass distributions for $B^- \to \Dz\pim$ and
$B^- \to \Dz\Km$ candidates, without and with use of information from
the RICH detector for $K/\pi$ separation.}{fig:BDK-RICH} 
\vspace*{-3ex}

The key to these new measurements is the excellent performance of the \Cleoiii\
tracking system and RICH detector.  The RICH provides very clean $K/\pi$
separation at the  momenta ($p \sim 2.5$ GeV/$c$) of the kaons and pions from
$\BtoKpi$, $\pi\pi$ and
$K\Kbar$ decay, \eg, within the fiducial volume (80\% of $4\pi$), the $K$
efficiency is 85\% with a 5\%  $\pi$ fake rate.  

The power of the RICH in reconstructing rare $B$ decays is illustrated in
\Fig{fig:BDK-RICH} which shows the $B$ mass peaks for $B^- \to \Dz\pim$ and
$B^- \to \Dz\Km$ candidates with and without using information from the RICH.  
Without the RICH, the signal for the Cabibbo suppressed $B^- \to \Dz\Km$ mode is 
overwhelmed with background from the Cabibbo favored $B^- \to \Dz\pim$ decay.
With the RICH, the $B^- \to \Dz\Km$ signal is almost free of background.
The previous CLEO measurement of $\calB(B^- \to \Dz\Km)$ required a very
sophisticated analysis, while the \Cleoiii\ measurement is simple and 
straight-forward.  The {\it preliminary} \Cleoiii\  branching
fraction
\Begeqn
\calB(B^- \to D^0 K^-) =
(3.8 \pm 1.3) \times 10^{-4}~~~
\Endeqn
agrees very well with the earlier CLEO result~\cite{cleo:BD0K}
\Begeqn
\calB(B^- \to D^0 K^-) =
(2.6 \pm 0.7) \times 10^{-4}.~~~
\Endeqn

\Begfigure{hbt}
\includegraphics[width=\Figwidth]{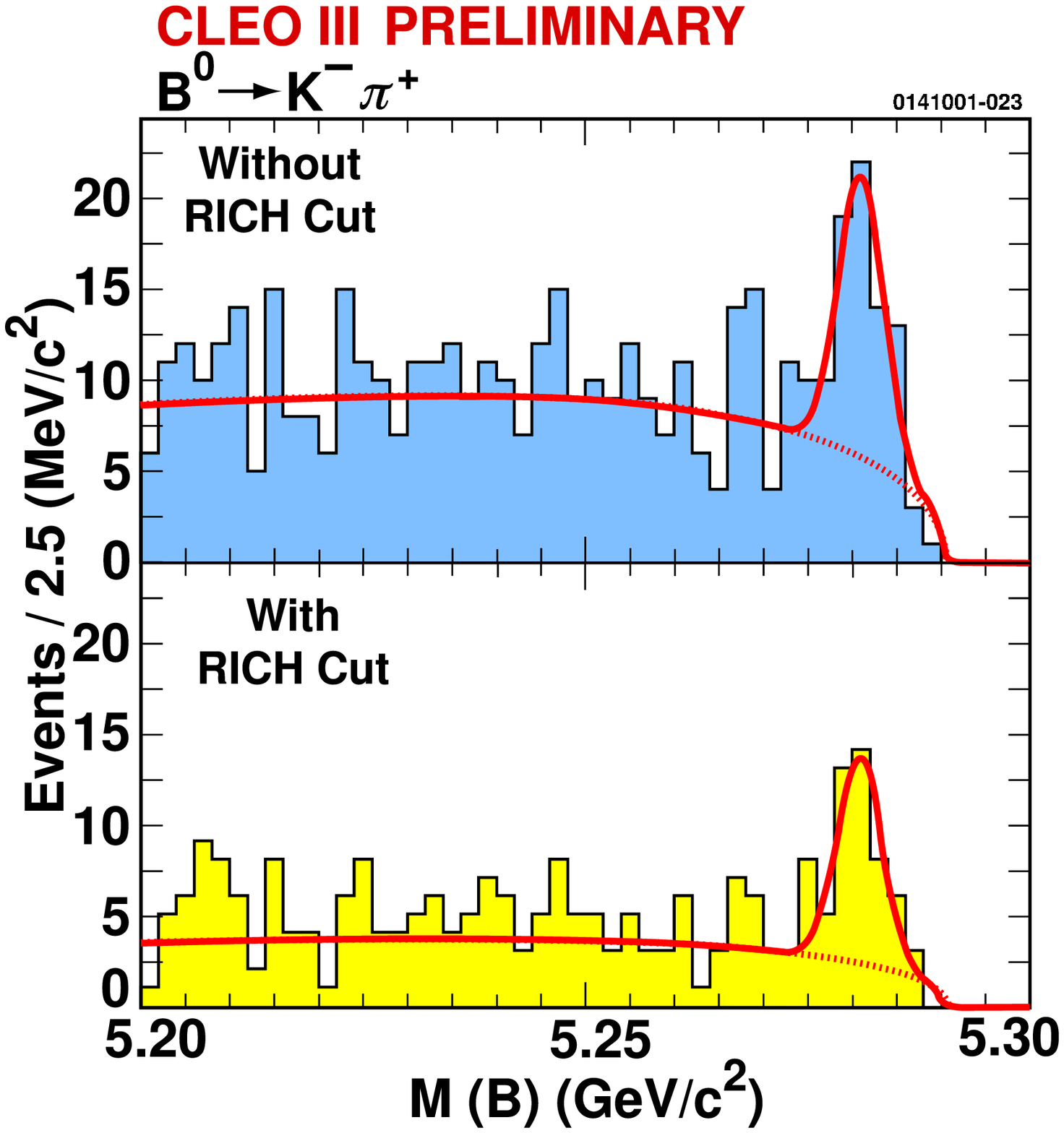}
\Endfigure{Reconstruction of $\Bz \to \Km\pip$ decays using the RICH
detector.  The top and bottom figures illustrates the $K\pi$ mass peaks without
and with using information from the 
RICH.}{fig:BzKpi-RICH}

\Fig{fig:BzKpi-RICH} shows that the RICH detector is also very effective in reducing
backgrounds

\twocolumn[\vspace*{-2ex}
\renewcommand{\Arraystretch}{1.5}
\Tablecaption{CLEO measurements of charmless two-body $B$ meson decays.  For
each decay mode, the first row is the {\it preliminary} \Cleoiii\ result and the
second row is the previously published result from \Cleoii\ and
\Cleoiiv\ data~\protect\cite{cleo:Brare00,cleo:Bpizpiz}. Upper limits (UL) are at the
90\%~CL.}\label{tab:brare:cleoiii}
\begin{center}
\Begtabular{|lclclc|} 
Mode & Efficiency & Yield & Significance & $\calB$ ($10^{-6}$) & UL ($10^{-6}$) \\
\hline
{$K^\pm\pi^\mp$}                        & 
{46\%}                                      & 
{$29.2_{-6.4}^{+7.1}$}                    & 
{5.4$\sigma$}                             & 
{\derr{18.6}{4.5}{4.1}{3.0}{3.4}}         & 
{}\\
{}                          & 
{48\%}                                      & 
{$80.2_{-11.0}^{+11.8}$}                  & 
{11.7$\sigma$~\,}                            & 
{$17.2^{+2.5}_{-2.4}\pm 1.2 $}            & 
\\
{$K^\pm\pi^0$}                            & 
{32\%}                                      & 
{\aerr{12.9}{6.5}{5.5}}                   & 
{3.8$\sigma$}                             & 
{\derr{13.1}{5.8}{4.9}{2.8}{2.9}}         & 
{}\\
 {}                            & 
 {38\%}                                      & 
 {$42.1_{-9.9}^{+10.9}$}                  & 
 {6.1$\sigma$}                             & 
 {$11.6^{+3.0}_{-2.7}{}^{+1.4}_{-1.3}$}    & 
 {}\\
{$K^0\pi^\pm$}                            & 
{12\%}                                      & 
{\aerr{14.8}{4.9}{4.1}}                    & 
{6.2$\sigma$}                             & 
{\derr{35.7}{12}{9.9}{5.4}{6.2}}             & 
{}\\
 {}                                        & 
 {14\%}                                      & 
 {$25.2_{-5.6}^{+6.4}$}                    & 
 {7.6$\sigma$}                             & 
 {$18.2^{+4.6}_{-4.0}\pm 1.6$}             & 
 {}\\
{$K^0\pi^0$}                              & 
~~~~{$\,8.5$\%}                                       & 
{\aerr{~\,3.0}{2.9}{2.5}}                    & 
{1.6$\sigma$}                             & 
{\derr{10.4}{10}{8.3}{2.9}{2.9}}       & 
{ }\\
 {}                                        & 
 {11\%}                                         & 
 {$16.1^{+5.9}_{-5.0}$}                    & 
 {$4.9\sigma$}                             & 
 {$14.6^{+5.9}_{-5.1}{}^{+2.4}_{-3.3}$}    & 
 {}\\
\hline
{$\pi^\pm\pi^\mp$}                       & 
{35\%}                                & 
{\aerr{~\,3.9}{1.5}{1.2}}                   & 
{2.2$\sigma$}                            & 
{\derr{~\,3.2}{3.3}{2.5}{1.0}{1.0}}         & 
{}\\
 {}                                       & 
 {48\%}                                        & 
 {$20.0^{+7.6}_{-6.5}$}                   & 
 {4.2$\sigma$}                            & 
 {$~\,4.3^{+1.6}_{-1.4}\pm0.5$}              & 
 {}\\
{$\pi^\pm\pi^0$}                         & 
{29\%}                                     & 
{\aerr{11.5}{5.6}{4.5}}                  & 
{3.4$\sigma$}                            & 
{\derr{11.7}{5.7}{4.6}{2.2}{2.4}}        & 
{}\\
 {}                                       & 
 {39\%}                                     & 
 {$21.3_{-8.5}^{+9.7}$}                   & 
 {3.2$\sigma$}                            & 
& 
{~~12.7}\\
{$\pi^0\pi^0$}                           & 
{29\%}                                     & 
{\aerr{~\,2.7}{2.4}{1.6}}                   & 
{2.9$\sigma$}                            & 
{}&
{11}\\
%
%
{}                           & 
{29\%}                                     & 
{\aerr{~\,6.2}{4.8}{3.7}}                    & 
{2.0$\sigma$}                            & 
{}&
~~~~{5.7}\\
%
\hline
{$K^\pm K^\mp$}                       & 
{36\%}                                  & 
{\aerr{~\,1.0}{2.4}{1.7}}                 & 
{0.6$\sigma$}                         & 
{}                                    & 
~~~~{4.5}\\
 {}                                    & 
 {48\%}                                  & 
 {$~\,0.7_{-0.7}^{+3.4}$}                 & 
 {0.0$\sigma$}                         & 
 {}                                    & 
~~~~{1.9}\\
{$K^0 K^\pm$}                       & 
{12\%}                                  & 
{\aerr{~\,0.5}{1.9}{1.1}}                 & 
{0.8$\sigma$}                         & 
{}                                    & 
{18}\\
 {}                       & 
 {14\%}                                  & 
 {$~\,1.4_{-1.3}^{+2.4}$}                 & 
 {1.1$\sigma$}                         & 
 {}                                    & 
~~~~{5.1}\\
{$K^0 \bar {K^0}$}                    & 
{13\%}                                   & 
{\aerr{~\,0.0}{0.5}{0.5}}                                   & 
{0.0$\sigma$}                                    & 
{}                                    & 
{13}\\
 {}                                     & 
 ~~{5\%}                                   & 
 ~{$\,0$}                 & 
 {0.0$\sigma$}                                     & 
 {}                                     & 
{17}\\
\Endtabular
\bigskip
\end{center}]

\Begfigure{b}
\ispect{$\bar{B}^0$}{$b$}{$\bar{d}$}
       {$c$}{$\bar{u}$}{$d$}{$\bar{d}$}{$D^{(*)0}$}{$\piz$}
\Endfigure{The internal spectator diagram for $\Bbar \to D^{(*)0} \piz$ decay.}
{fig:BDzpiz}

\noindent  and improving the signal to noise ratio in $\Bz\to\Kp\pim$
decays.

The other techniques used in reconstructing these decays are similar to
those used in the previous CLEO
analyses~\cite{cleo:Brare98,cleo:Brare00,cleo:Bpizpiz}. The {\it preliminary}
results from \Cleoiii\ data are compared to the previous CLEO measurements in
\Tab{tab:brare:cleoiii}.  In all cases the \Cleoiii\
results agree very
well with the earlier CLEO measurements.  The final results from the full
\Cleoiii\ data sample will be combined with the earlier CLEO measurements. 

\section{Other Recent CLEO Results}

In this section I briefly mention a few other recent CLEO results.

Internal spectator diagrams (see \Fig{fig:BDzpiz}) in $\Bbar$ decay are
processes in which the $W$ from  $b \to c(u)$ decay produces a
$q\bar{q}'$ pair that hadronizes with $c(u)$ quark and the antiquark from the
$\Bbar$.  These decays are suppressed because the color of the  $q\bar{q}'$
quarks does not automatically match the colors in the quarks from the $\Bbar$
fragment.  So far the only color suppressed $\Bbar$ decays that have been
observed have charmonium  in the final state, \eg, $\Bzbar \to \Jpsi \Kzbar$. 
CLEO has now observed~\cite{cleo:Dzpiz} the first color suppressed decays
without charmonium in the final state, 
$\Bbar^0 \to \Dz\piz$ and $\Bbar^0 \to \Dstarz\piz$.  

The branching fractions measured for these decays are
\Begeqnarray
\calB(\Bbar^0 \to \Dz~\piz) &=& (\aerr{2.74}{0.36}{0.32} \pm 0.55) \nonumber \\
&\times& 10^{-4}~~~\mathrm{and} \\
\calB(\Bbar^0 \to \Dstarz\piz) &=& (\aerr{2.20}{0.59}{0.52} \pm 0.79) \nonumber \\
&\times& 10^{-4}.
\Endeqnarray
The statistical significances of these signals are $12\sigma$ and $5.9\sigma$
for $\Dz\piz$ and $\Dstarz\piz$, respectively.

CLEO studies of $B \to K\pi$ and $B \to \pi\pi$ decays demonstrated that
gluonic penguin diagrams are important in $B$ decay (see \Sec{sec:brare}).
However, final states such as $B \to \phi K$ and $B \to \phi K^*$ play
a special role since they cannot be produced at a significant rate by any other decay
mechanism.  Earlier, CLEO reported~\cite{cleo:BphiK} the first significant
measurement of some of these decays. The branching fractions,
\Begeqnarray
\calB(\Bm \to \phi \Km) &=& (\aerr{5.5}{2.1}{1.8}\pm 0.6)
\times\! 10^{-6} \\
\mathrm{and~~~~~~~~~~~~~~~} &~&  \nonumber \\
\calB(\Bz \to \phi K^{*0}) &=& (\derr{11.5}{4.5}{3.7}{1.8}{1.7})\,
\times\! 10^{-6},
\Endeqnarray
were measured at significance levels of $5.4\sigma$ and $5.1\sigma$, respectively.
These branching fractions are well within the rather large ranges predicted by
theoretical models (see Ref.~\cite{cleo:BphiK}).
Indications of the other charge modes, $\phi\Kz$ and $\phi K^{*-}$, were
observed at the $\sim 3\sigma$ level. 

The FCNC decays $B \to K\ell^+\ell^-$ and $B \to K^*\ell^+\ell^-$
are another window on effects from possible New Physics in radiative penguin loops, 
since the $\gamma$ in $\ell^+\ell^-$ decays is virtual.  In particular, 
$B \to K\gamma$ is forbidden by angular momentum conservation, while  
$B \to K\ell^+\ell^-$ is allowed. So far these exclusive decays have not been
observed.  CLEO recently reported~\cite{cleo:BKll} improved upper limits
\Begeqnarray
\calB(B \to K~\ell^+\ell^-) &<& 1.7 \times 10^{-6}~~~\mathrm{and} \\
\calB(B \to K^*\ell^+\ell^-) &<& 3.3 \times 10^{-6}
\Endeqnarray
at the 90\% CL.  (For the $\calB(B \to K^*\ell^+\ell^-)$ limit, the dilepton
mass range is
$m_{\ell\ell} > 0.5$ GeV.) The limit, $\calB(B \to K^{(*)}~\ell^+\ell^-) <1.5 \times
10^{-6}$, obtained for the weighted average of these decays, is not very far
above the theoretical prediction~\cite{alietal}: $1.0 \times 10^{-6}$.

The decay $B^+ \to \Dstarp \Kzs$ should proceed via an annihilation diagram
in which the $W^+$ from the annihilation of the $\bar{b}$ and $u$ quarks produces
a $c\bar{s}$ pair which hadronizes to $\Dstarp \Kzs$. No reliable theoretical
prediction for the rate of this decay exists.  We searched~\cite{cleo:BD*K0} for
this decay and determined an upper limit at the 90\% CL:
\Begeqn
\calB(B^+ \to \Dstarp \Kzs) < 9.5 \times 10^{-5}
\Endeqn

\section{CLEO-c and CESR-c}

\Cleoc\ is a focused program of measurements and searches in $e^+e^-$ collisions in
the the $\sqrt{s}=3-5$ GeV energy region~\cite{YB}.  Topics to be studied include:
\Begitem
\item Precision -- $\calO(1\%)$ -- charm measurements:
absolute charm branching fractions,
the decay constants $\fdp$ and $\fds$, 
semileptonic decay form factors, and the CKM matrix elements
$\Vcd$ and $\Vcs$
\item Searches for New Physics in the charm sector:
$CP$ violation in $D$ decay,
$D\Dbar$ mixing without doubly suppressed Cabibbo decay, and
rare $D$ decays
\item $\tau$ studies: precision measurements and searches for New Physics
\item QCD studies:
$c\bar{c}$ spectroscopy,
searches for glue-rich exotic states (glueballs and hybrids), and
measurements of $R$ 
(direct between 3 and 5 GeV, and
indirect using initial state radiation between 1 and 3 GeV)
\Enditem
The \Cleoiii\ detector described above is a crucial element of this program.
Its capabilities and performance are substantially beyond those of other
detectors that have operated in the charm threshold region.

Testing Lattice QCD (LQCD) calculations with precision measurements is a major
emphasis of the \Cleoc\ program. Theoretical analysis of strongly-coupled,
nonperturbative quantum field theories remains one of the foremost challenges in
modern physics. Experimental progress in flavor physics (\eg, determining the CKM
matrix elements $\Vcb$, $\Vub$, and $\Vtd$) is frequently limited by knowledge of
nonperturbative QCD effects, (\eg, decay constants and semileptonic form factors). 
(One sort of approach to reducing theoretical uncertainties in determining $\Vcb$ and
$\Vub$ was already described in Secs.~\ref{sec:Vcb-mx} and \ref{sec:Vub}.)
In the last decade several technical problems in LQCD have been identified and
overcome, and substantially improved algorithms have been developed. LQCD
theorists are now poised to move from
$\calO(15\%)$ precision to
$\calO(1\%)$ precision in calculating many important parameters that can be measured
experimentally or are needed to interpret experimental measurements, including:\break
\Begitem\vspace*{-3ex}
\item masses, leptonic widths, EM transition form factors, and mixing amplitudes of
$c\bar{c}$ and $b\bar{b}$ bound states; and
\item masses, decay constants, and semileptonic decay form factors of $D$ and $B$ mesons.
\Enditem
\Cleoc\ will provide data in the charm sector to motivate and validate many of these
calculations.  This will help to establish a comprehensive mastery of
nonperturbative QCD, and enhance confidence in LQCD calculations in the
beauty sector.

The \Cleoc\ program is based on a four-year run plan, where the first year is spent on
the $\Upsilon$ resonances while constructing hardware for CESR improvements.
We expect to accumulate the following data samples:
\begin{list}{$\bullet$}{
\settowidth{\leftmargini}{2005~~}
\settowidth{\leftmargin}{2005~~}
\settowidth{\labelwidth}{2005}
\setlength{\labelsep}{\leftmargin}
\addtolength{\labelsep}{-\labelwidth}
\Itemspace}
\item[2002] $\gtsim 1$ \fbinv\  at
each of the
$\Upsilon(1S)$,
$\Upsilon(2S)$, $\Upsilon(3S)$ resonances\\ (10-20 times the
existing world's data)
\item[2003] 3 \fbinv\ at the $\psi(3770)$ -- 30 M $D\Dbar$ events and 6 M tagged $D$
decays\\ (310 times the MARK III data)
\item[2004] 3 \fbinv\ at $\sqrt{s} \sim 4.1$ GeV -- 1.5 M $\Ds\Dsbar$ events and
0.3 M tagged $\Ds$ decays\\
(480 times the MARK III data and
130 times the BES II data)
\item[2005] 1 \fbinv\ at the $\Jpsi$ --  1 G $\Jpsi$ decays\\
(170 times the MARK III data and 20 times the BES II data)
\end{list}
Detailed Monte Carlo simulations show that we will be able to measure the
charm reference branching fractions, decay constants, slopes of semileptonic form
factors, and CKM matrix elements -- all with $\calO(1\%)$ precision.  

Goals of the run on the $\Upsilon$ 
bound states include searches for the ``missing $b\bar{b}$ states'' 
-- \eg, $^1S_0$ ($\eta_b$, $\ldots$) and $^1P_1$ ($h_b$, $\ldots$) --
and accurate measurements of  $\Gamma_{ee}$'s, transition rates, and hyperfine
splittings.   Most of the quantities that will be measured in this program 
can be used to validate precise LQCD calculations.

The 1 G $\Jpsi$ events will be an extremely rich
source of data for glueball searches.  
The very controversial $f_J(2220)$ is an excellent example of the enormous reach of
this program.  Using the values of
$\calB(\Jpsi \to \gamma f_J)\calB(f_J \to Y\bar{Y}$) measured by BES~\cite{bes:fj} we
expect peaks with 23,000, 13,000, and 15,600 events would be observed in the
$f_J(2220)$ decay channels $\pip\pim$,
$\piz\piz$, and $\Kp\Km$, respectively.  All of these signals would stand out well
above reasonable estimates of backgrounds. 

We have just completed the installation of superconducting interaction region
quadrupoles that will allow us to operate CESR over the energy range from charm
threshold to above the $\Upsilon(4S)$.  In the
$\Upsilon$ region, synchrotron radiation damping reduces the size of beams in CESR
and is a crucial factor for achieving high luminosity.  This damping will be much
less at lower energies in the charm threshold region, and that would substantially
reduce  luminosity.  Much of this luminosity loss can be recovered by installing
wiggler magnets (magnets with alternating magnetic field directions) to increase
synchrotron radiation.  We plan to use superferric wiggler magnets (Fe poles and
superconducting coils) and we have already constructed a three-pole prototype.  The
anticipated luminosity will still be below that achieved in the
$\Upsilon$ region, and will increase with energy, ranging from $0.2\times
10^{33}$~\Lunits\ to $0.4\times 10^{33}$~\Lunits\ in the energy region between
3.1 and 4.1 GeV, and rising to
$\gtsim 1\times 10^{33}$~\Lunits\ in the
$\Upsilon$ region. These wiggler magnets are the only substantial CESR hardware
upgrade required for the
\Cleoc\ program, and -- not entirely incidentally -- they are also excellent prototypes
for the wiggler magnets that would be needed in linear collider damping rings.

\vspace*{-2ex}

\section{Summary and Conclusions}

\vspace*{-1ex}

We report new results based on the full \Cleoii\ and \Cleoiiv\ data samples. 
For $\btosgamma$ decays we find
\Begeqnarray
\calB(\btosgamma) &=& 
(3.21 \pm 0.43 \pm 0.27 ^{+0.18}_{-0.10}) \nonumber \\ 
&\times& 10^{-4}
\Endeqnarray
where the first error is statistical, the second is systematic, and the third
is from theoretical corrections.  We substantially reduced the theoretical 
uncertainties that occurred in our earlier measurement~\cite{cleo:bsgamma1} by
including nearly all of the photon energy spectrum. We measured  $\Vcb$ using
$\BtoXclnu$ hadronic mass moments and
$\btosgamma$ energy moments, again with substantially reduced theoretical
uncertainties.  The result is
\Begeqn
\Vcb = (40.4 \pm 0.9 \pm 0.5 \pm 0.8) \times 10^{-3}
\Endeqn
where the errors are due to uncertainties in moments, $\Gamma^c_{SL}$, and 
theory, in that order. We measured $\Vub$ using the $\btosgamma$ spectrum to
determine the fraction of the $\BtoXulnu$ lepton momentum spectrum in the
momentum interval used.   The {\it preliminary} result is
\Begeqn
V_{ub} = (4.09 \pm 0.14 \pm 0.66) \times\! 10^{-3}
\Endeqn
where the first error is statistical and the second is systematic.  
Again, theoretical uncertainties are substantially less than those in
previous measurements.

We report a {\it preliminary} new measurement of $\Vcb$ from both
$\BztoDstarplnu$  and
$\BmtoDstarzlnu$ decays.  The result is
\Begeqn
\Vcb = (46.2 \pm 1.4 \pm 2.0 \pm 2.1) \times 10^{-3}
\Endeqn
where the errors are statistical, systematic, and theoretical, respectively.

We present the first {\it preliminary} results from \Cleoiii\ data --
measurements of and upper limits for
$\calB(B \to K\pi)$, $\calB(B \to \pi\pi)$, and $\calB(B \to K\Kbar)$.

Finally, we are embarking on a new program of operating CESR and CLEO at
the $\Upsilon$ bound states and in the charm threshold region.  This program
will yield: precision measurements of $\Upsilon$ parameters, searches for missing
$b\bar{b}$ states, precision measurements in the charm and tau sectors,
searches for New Physics in charm and tau decays, and definitive searches for
low-lying glueball states.  This diverse program will be unified by collaboration
with Lattice QCD theorists who will use the results to validate their calculations
and gain confidence for their utilization in the $b$ quark sector.

\vspace*{-1ex}

\section*{Acknowledgments}

I am delighted to express my appreciation for my CESR and CLEO
colleagues whose effort provided the results described here, and for
the contributions of our theoretical colleagues to these measurements.  Our research was
supported by the NSF and DOE. I appreciate the hospitality of the DESY Hamburg
laboratory where I prepared this report for the conference proceedings.  Fabio Anulli
provided invaluable assistance in his role as Scientific Secretary.  Finally, I want to
thank Paolo Franzini, Juliet Lee-Franzini, and the entire staff of Lepton-Photon 2001
for the delightful conference that they organized.

\def\Journal#1#2#3#4{{#1} {\bf #2}, #3 (#4)}
\def\Report#1#2#3{#1 Report No.\ #2 (#3)}
\def\Journalmore#1#2#3{{\bf #1}, #2 (#3)}

\def\CLEO{CLEO Collaboration}
\def\EPJC{{\it E.\ Phys J.} C}
\def\NCA{\it Nuovo Cimento}
\def\NIM{\it Nucl.\ Instrum.\ Methods}
\def\NIMA{{\it Nucl.\ Instrum.\ Methods} A}
\def\NPB{{\it Nucl.\ Phys.\ } B}
\def\PLB{{\it Phys.\ Lett.\ } B}
\def\PRL{\it Phys.\ Rev.\ Lett.\ }
\def\PRD{{\it Phys.\  Rev.\ } D}
\def\ZPC{{\it Z.\ Phys.\ } C}

\def\etal{{\it et al.}}


\begin{thebibliography}{99}
\bibitem{lep:hfwg}LEP data are summarized and corrected to common daughter
branching fractions by the LEP Heavy Flavor Working Group
\end{thebibliography}
\end{document}